\documentclass[namedreferences]{solarphysics}

\usepackage[hyperref,optionalrh,natbib]{spr-sola-addons} % For Solar Physics 
\usepackage{multirow}
\usepackage{graphicx}        % For eps figures, newer & more powerfull
\usepackage{color}           % For color text: \color command
\usepackage{breakurl}        % For breaking URLs easily trough lines
\begin{document}

\begin{article}

\begin{opening}

\title{Diagnostics of the Coronal Hole and the adjacent Quiet Sun  by {\it The Hinode/EUV Imaging Spectrometer} (EIS)}

\author{P.~\surname{Kayshap}$^{1}$\sep
        D.~\surname{Banerjee}$^{2,3}$\sep
        A. K.~\surname{Srivastava}$^{4}$      
       }
\runningauthor{Kayshap et al.}
\runningtitle{Comparison Between Coronal Holes and Quiet-Sun}

   \institute{$^{1}$ Aryabhatta Research Institute of Observational Sciences (ARIES), Manora Peak, Nainital-263002, India\\
                     email: \url{pradeep.kashyap@aries.res.in}\\ 
              $^{2}$ Indian Institute of Astrophysics (IIA), Bangalore-560034, India\\
                     email: \url{dipu@iiap.res.in}\\
                 $^{3}$ CESSI, Indian Institute of Science Education and Research, Kolkata, Mohanpur - 741252, India\\
                     email: \url{dipu@iiap.res.in}\\     
              $^{4}$ Department of Physics, Indian Institute of Technology (Banaras Hindu University), Varanasi-221005\\
                     email: \url{asrivastava.aap@iitbhu.ac.in}\\ 
             }

\begin{abstract}
A comparison between a Coronal Hole (CH) and the adjacent Quiet-Sun (QS) has been performed using the spectroscopic diagnostics  of {\it Hinode/ the EUV Imaging Spectrometer} (EIS). Coronal funnels play an important role in the formation and propagation of the nascent fast solar wind.  Applying Gaussian fitting procedures to the observed line profiles, Doppler velocity, intensity, line width (FWHM) and electron density have been estimated over CH and adjacent QS region of a North Polar Coronal Hole (NPCH). The aim of this study is to identify the coronal funnels based on spectral signatures. Excess width regions (excess FWHM above a threshold level) have been identified in QS and CH. The plasma flow inversion (average red-shifts changing to blue-shifts at a specific height) in CH  and excess width regions of QS take place at $\sim$ 5.01$\times$10$^{5}$ K. % while average width regions of QS do not show conclusive plasma flow inversion temperature. %The properties of both type regions of CH are similar to the excess width regions of QS. 
Furthermore, high density concentration in excess width regions of QS provides an indication that these regions are %the CH counterpart ({\it i.e.}, 
the footprints of coronal funnels. We have also found that non-thermal velocities of CH are higher in comparison to QS confirming that the CHs are the source regions of the fast solar wind.  Doppler and non-thermal velocities as recorded by different temperature lines have been also compared with previously published results. As we go  from lower to upper solar atmosphere, down-flows are dominated in lower atmosphere while coronal lines are dominated by up-flows with a maximum value of $\sim$ 10-12 km s$^{-1}$ in QS. Non-thermal velocity increases first but after Log T$_{e}$ = 5.47 it decreases further in QS. This trend can be interpreted as a signature of the dissipation of Alfv{\'e}n waves, while increasing trend as reported earlier may attribute to the signature of the growth of Alfv{\'e}n waves at lower heights. Predominance of occurrence of nano-flares around O~{\sc vi} formation temperature could also explain the non-thermal velocity trend.   
\end{abstract}
\end{opening}
%-------------------------------------------------
\section{Introduction}
     \label{S-Introduction} 
Coronal Holes (CHs; \citealt{Wald1975}) are the regions on the Sun, which appear dark in comparison to the quiet-Sun (QS) because they emit less in UV $\&$ X-rays and maintained at a lower temperature than the surroundings. The different magnetic structures of QS and CH are responsible for their different appearance in coronal lines \citep{Wmann2004}. CHs are dominated by coronal funnels, which originate from the adjoining locations of super-granule boundaries at the photosphere and expand abruptly towards higher atmosphere \citep{Gabriel1976, Axford1997, Marsch1997, Hackenberg1999}. QS regions are dominated by closed magnetic filed lines. Although, QS has also the coronal funnels with less expansion. The filling factors of these coronal funnels are less in QS compared to CHs. Coronal funnels play an important role in the formation as well as propagation of nascent fast solar wind in the CHs \citep{Tu2005, He2008,Tian2010}. Spectroscopic analysis is important  to understand the dynamics of the QS as well as CH. The Doppler velocity, which provides important information to any model of solar coronal heating, has a tendency to invert from pre-dominant red-shifts ({\it i.e.}, downflows) to blue-shifts ({\it i.e.}, upflows) as we go higher from chromosphere to corona in CHs \citep{Warren1997, Hassler1999, PJ1999, Peter1999, Xia2003} as well as in QS \citep{Chae1998, PJ1999, Teriaca1999, Neda2011}. Several possible mechanisms have been proposed  to explain the observed red-shifts of the solar transition region (TR) and coronal lines, {\it e.g.}, return of spicular material \citep{Pneuman1978,Athay1984}, downward propagating acoustic waves \citep{Hansteen1993, Hansteen1997,Teriaca1999}. The turnover temperature from red-shift to blue-shift is a crucial observational parameter for future modelling. %to understand the observed pattern of Doppler velocity from chromosphere to corona. %In spite of the several explanations of the observed Doppler velocity pattern, the nature of the Doppler velocity is not fully understood. \\ 
%After the Solar Ultraviolet Meaursement of Emitted Radiation (SUMER)/SoHO era, In the CHs Ne~{\sc vii} 770~\AA\, formed in the upper TR, is blue-shifted (e.g.,\cite{Dam1999,PJ1999,Wilhelm2000, Xia2003, Aioz2005, Mcintosh2007,Tian2008a, Tian2008c}), which was suggested that these blue-shifted patches at upper TR are associated with coronal funnels and interpreated as the initial outflow of the solar wind (e.g., \citealt{Tu2005, He2008}). Recently, on the basis of coronal emission lines from Hinode/EIS spctra \cite{Tian2010} have shown, which is the extanison of these work, the nacent fast solar wind originates in the coronal funnels.\\

Full-width-at-half-maximum (FWHM) is also an important parameter to understand the dynamics of QS and CH. Some observed solar Extreme-ultraviolet (EUV) and Far-ltraviolet (FUV) spectral line profiles are found to be broader than those expected from thermal broadening  \citep{Boland1975, Doschek1976, Mariska1978, Banerjee1998}. These excess widths, after subtracting thermal and instrumental widths from the observed widths, are called as non-thermal widths. Non-thermal velocities as derived from these widths reveal the presence of non-thermal motions/unresolved flows or the presence of waves in the solar atmosphere. Early studies show that the non-thermal velocities  increase within a narrow range of temperature (from chromosphere up to TR temperatures) in the QS near disk centre \citep{Doschek1976, Mariska1978, Dere1993}). After the launch of Solar Ultraviolet Measurements of Emitted Radiation (SUMER) on board SoHO, the non-thermal velocity has been investigated for a quite large range of solar atmosphere ({\it i.e.}, 10$^{4}$ K to 2$\times$10$^{6}$ K). It is found that initially non-thermal velocity increases but after a certain temperature decreases corresponding to coronal heights in QS as well as active regions \citep{Chae1998a, Teriaca1999, Doschek2000, Peter2001, Pat2006, Doschek2007}). Several mechanisms have been proposed to interpret the pattern of non-thermal velocity through the solar atmosphere, {\it e.g.}, MHD wave models or magnetic reconnection generated turbulence \citep{Doschek1976, Mariska1978}, unresolved laminar flows, waves and turbulent flows \citep{Chae1998}, Alfv{\'e}n waves \citep{Peter2001}, {\it etc}. Recently, \cite{Coyner2011} have shown that the non-thermal velocities represent strong distribution from 19 km s$^{-1}$ to 22 km s$^{-1}$ in the various parts of the solar atmosphere. %Although, several possible mechanisms ({\it e.g.}, magnetic reconnection, unresolved laminar flows, acoustic wave modes, MHD wave modes, Alfv{\'e}n waves, etc.) have been suggested for the non-thermal broadening in the solar atmosphere but still we are looking for the satisfactory theory, which is a subject for further investigation.\\

In the present study, we have performed a comparison between QS and CHs on the basis of various spectral parameters ({\it e.g.}, Doppler velocity, non-thermal width, {\it etc}.). On the basis of the variations of these parameters and density contrast, we have located the most probable regions of coronal funnels in QS. Variation of the non-thermal velocity and average Doppler velocity with temperature have also been investigated in QS by including some previous relevant results with our estimated values. %The used data-set in the present work has been already used in two previously published works. First work shows the presence of Alf{\'e}n waves in the polar coornal hole \citep{Banerjee2009} while the second work represents the origin of the nascent fast solar wind in polar coronal holes\citep{Tian2010}. 
The present work is organized as follows. In Section~\ref{S-data-analysis}, we describe the details about the observation and data reduction. Results related to the comparison between QS and CH as well as identification of the coronal funnels are presented in Section~\ref{S-QS-CH}. Section~\ref{S-T-DV_NTV} outlines the temperature dependent behaviour of non-thermal velocity in QS and CH. Discussion and Conclusions are presented in
the last section.% Section~\ref{S-Discussion}, and Conclusions are mentioned in the last Section.
\section{Observations and Data Reduction} %%%%%%%%%%%%%%%%%%%%%%%%%%%%%%%%%%%%%%%%
\begin{figure}[ht]
   \centerline{\includegraphics[scale=0.8,clip=true]{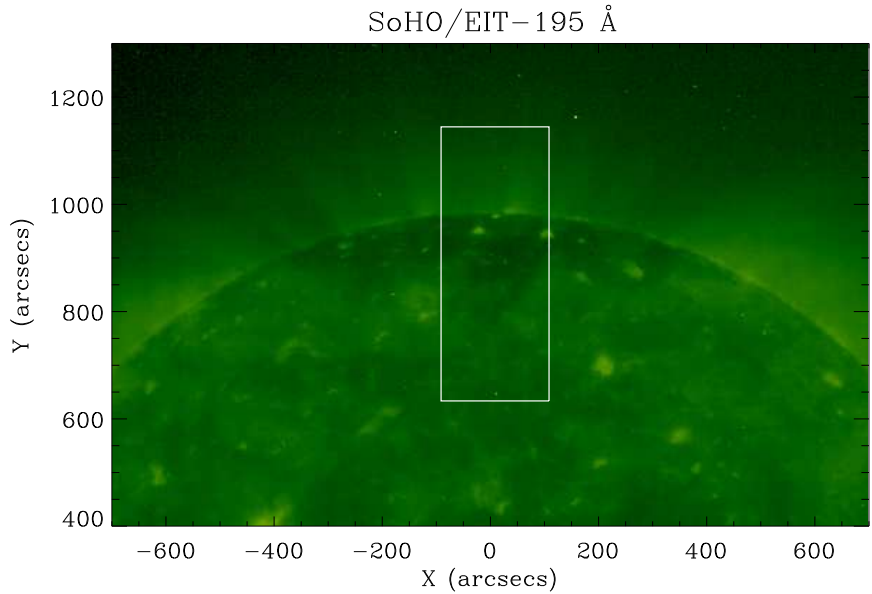}
              }
              \caption{White rectangular box marks the FOV of the observed region of {\it Hinode}/EIS, over-plotted on the SoHO/EIT 195~\AA\ image. This North Polar Coronal Hole (NPCH) was observed by {\it Hinode}/EIS from 14:13 UT to 18:17 UT on 10 October 2007.} 
\label{fig:fig1}
\end{figure}

      \label{S-data-analysis}      
{\it Extreme Ultraviolet Imaging Spectrometer} (EIS) on board {\it Hinode} spacecraft is a normal-incidence EUV spectrometer. EIS has spatial resolution of $\sim$ 2-arcsecond and high spectral resolution of 0.0223 \AA\ per pixel. EIS observes high resolution spectra in short wavelength band ({\it i.e.}, 170 - 211 \AA) and long wavelength band ({\it i.e.}, 246 - 292 \AA). Four types of slits/slots ({\it e.g.},1-,2-,40-,256-arcsecond) are available in the EIS observations \citep{Culhane2007}. A North Polar Coronal Hole (NPCH) observation captured on 10 October 2007 by {\it Hinode}/EIS, has been used in the present work. The observed region (FOV) is marked as a white rectangular box on the SoHO/EIT 195~\AA\ image ({\it cf.}, Figure~\ref{fig:fig1}). 
%+++++++++++++++++++Figure 1++++++++++++++++++++++++++++++++++++++++++++++++++++++++++++++++
%+++++++++++++++++++++++++++++++++++++++++++++++++++++++++++++++++++++++++++++++++++++++++++++
NPCH was observed using the 2-arcsecond slit in the raster  mode for more than four hours ({\it i.e.}, from 14:03 UT to 18:17 UT) by capturing 101 exposures with 155 s exposure time on each scanning step. This very long exposure ({\it i.e.}, 155 second) enhances the count statics, therefore, this data set is  reliable for deriving  various line parameters (e.g., Doppler velocity, FWHM, etc.) from the line profiles and also ideal for density diagnostics. Standard Hinode/EIS routine eis$\_$prep.pro, which is available in SolarSoft (SSW) package, has been used for the calibration of raw data. This routine converts original raw data into physical data after processing various necessary steps, {\it e.g.}, subtraction of the dark current, removal of cosmic rays and  hot pixels and radiometric calibration, {\it etc.} Four spectral lines have been used in the present analysis as listed in Table~1.
%++++++++++++++++++++Figure 2++++++++++++++++++++++++++++++++++++++++++++++++++++++
\begin{figure}
\centerline{\hspace*{-0.025\textwidth}
            \includegraphics[trim= 0.5cm 0.5cm 0.5cm 0.0cm,height=0.18\textheight, width=0.5\textwidth]{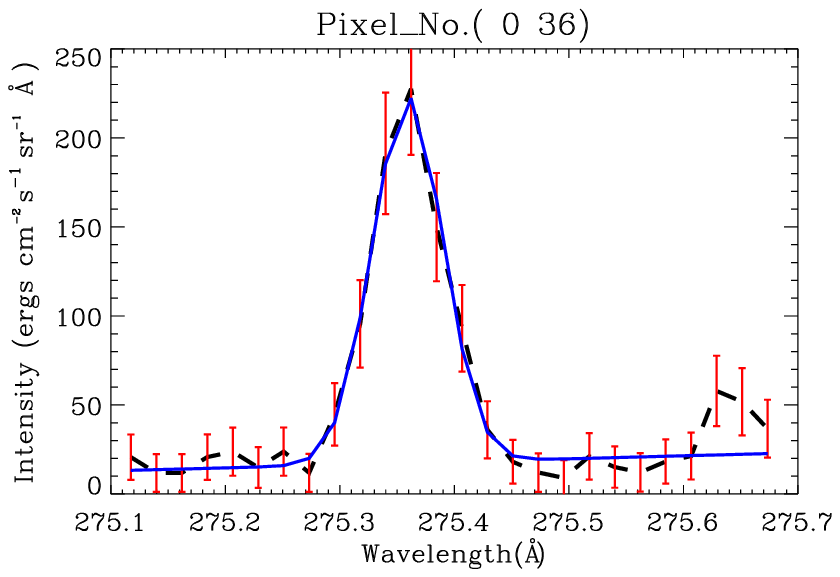}
            \includegraphics[trim= 0.0cm 0.5cm 0.5cm 0.0cm,height=0.18\textheight, width=0.5\textwidth]{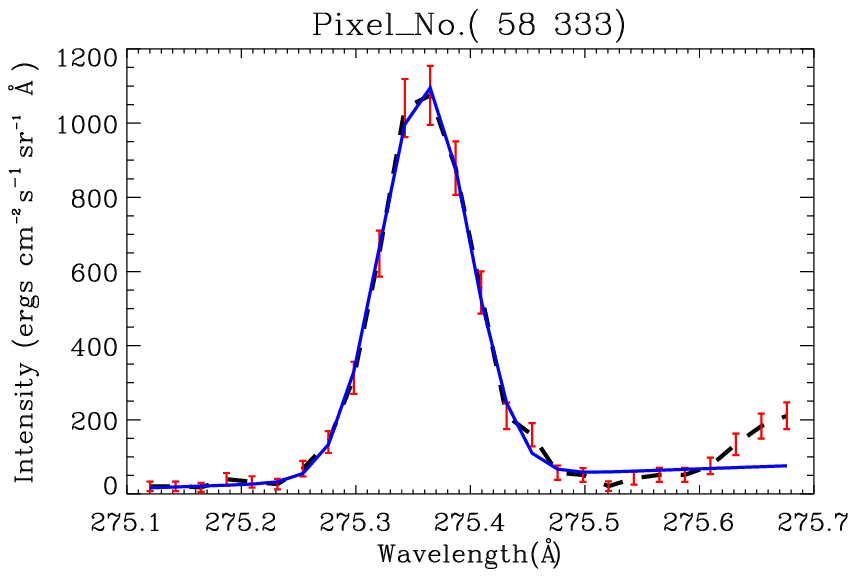}
            }
\vspace{0.4cm}
\centerline{\hspace*{-0.025\textwidth}
            \includegraphics[trim= 0.0cm 0.5cm 0.5cm 0.0cm,height=0.18\textheight, width=0.5\textwidth]{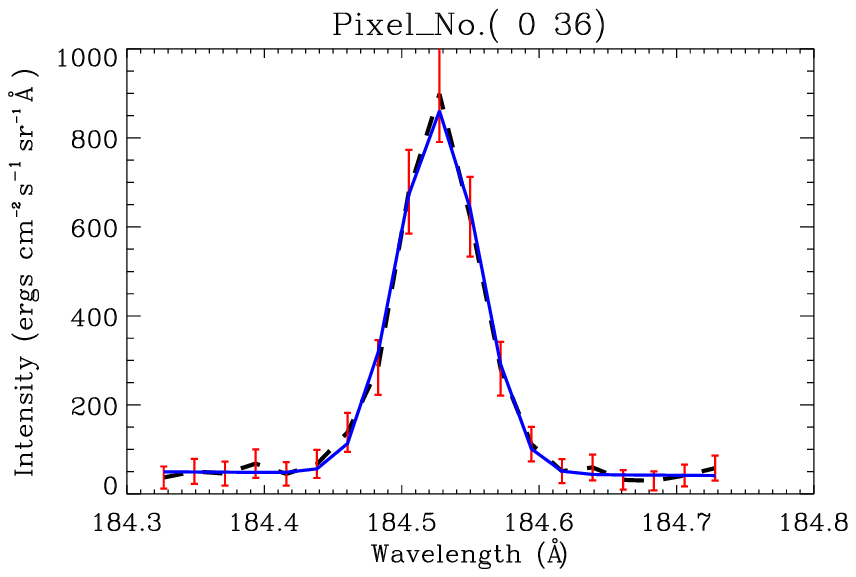}
            \includegraphics[trim= 0.0cm 0.5cm 0.5cm 0.0cm,height=0.18\textheight, width=0.5\textwidth]{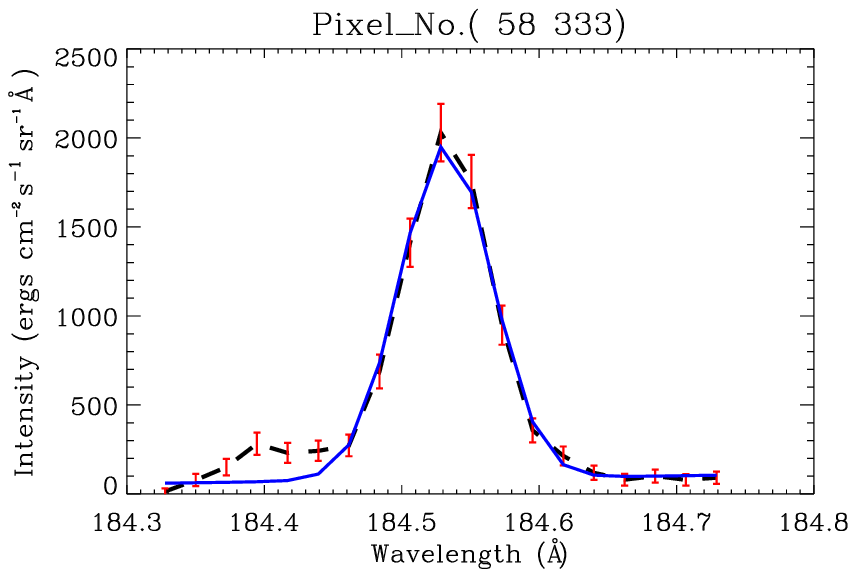}
}
\vspace{0.4cm}
\centerline{\hspace*{-0.025\textwidth}
           \includegraphics[trim= 0.0cm 0.5cm 0.5cm 0.0cm,height=0.18\textheight, width=0.5\textwidth]{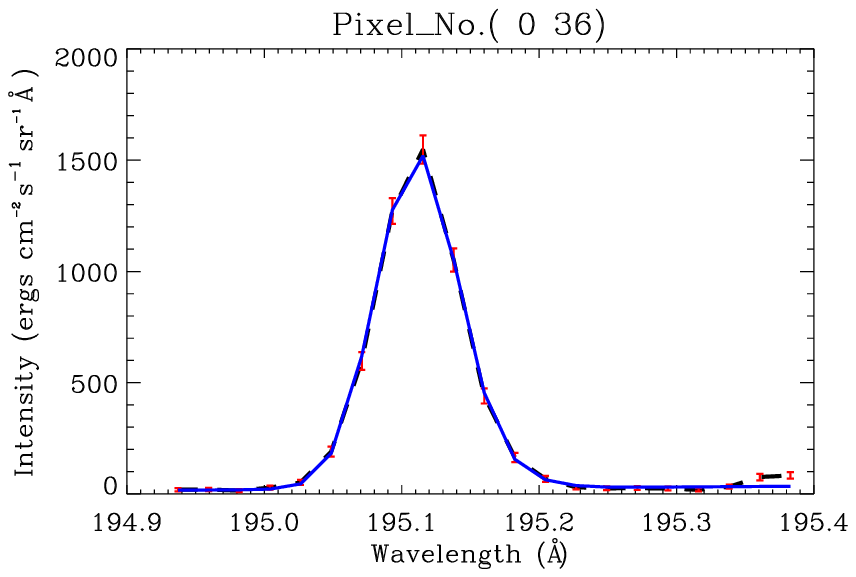}
           \includegraphics[trim= 0.0cm 0.5cm 0.5cm 0.0cm,height=0.18\textheight, width=0.5\textwidth]{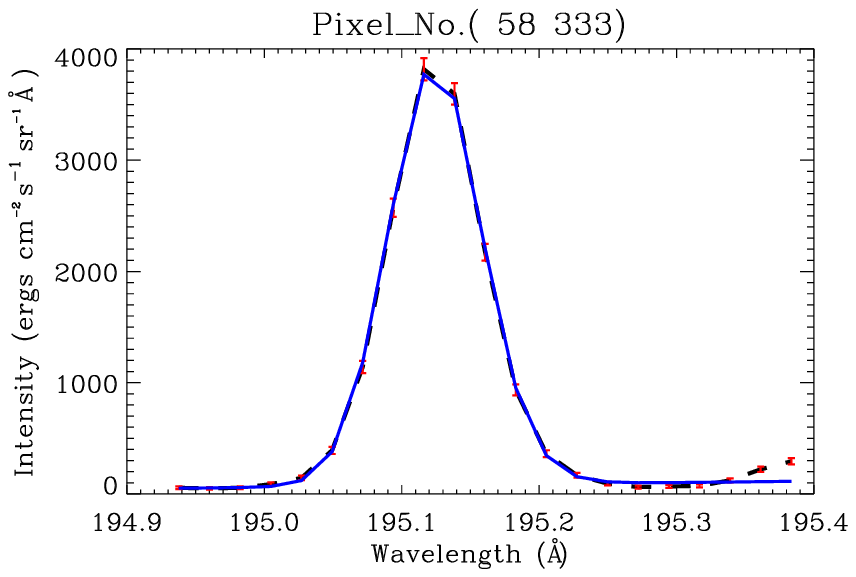}
}
\vspace{0.4cm}
\centerline{\hspace*{-0.025\textwidth}
            \includegraphics[trim= 0.0cm 0.5cm 0.5cm 0.0cm,height=0.18\textheight, width=0.5\textwidth]{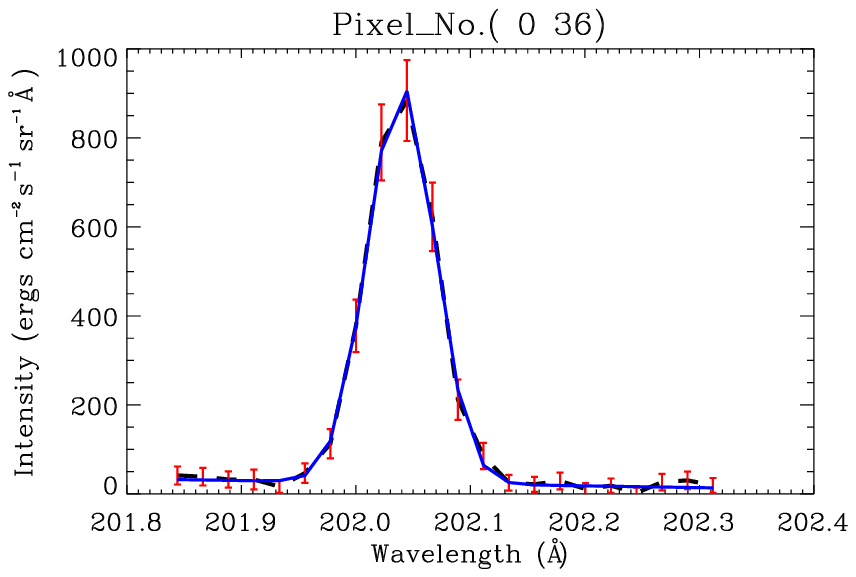}
            \includegraphics[trim= 0.0cm 0.5cm 0.5cm 0.0cm,height=0.18\textheight, width=0.5\textwidth]{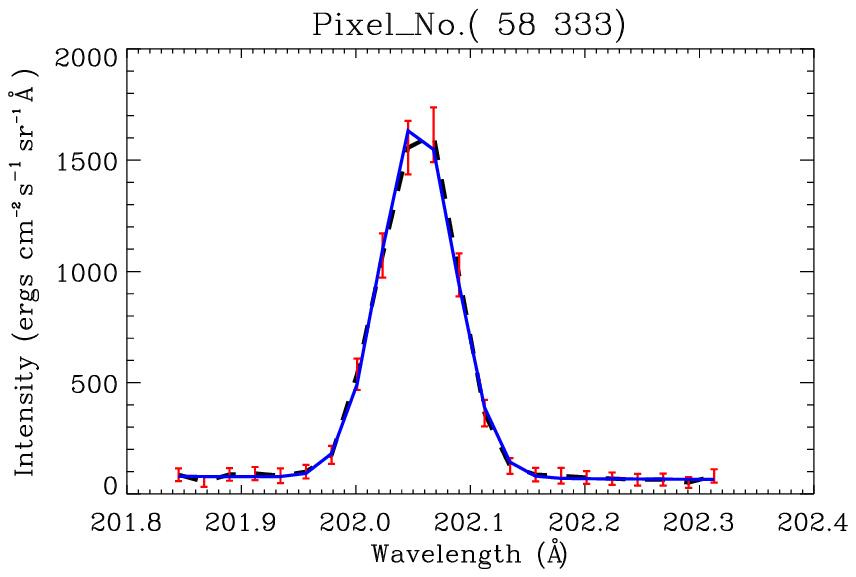}
}
\caption{ Two samples of Gaussian fitting (shown by blue solid line) in Si~{\sc vii} 275.35~\AA~(top-row), Fe~{\sc x} 184.54~\AA~(second-row), Fe~{\sc xii} 195.120~\AA~(third-row) and Fe~{\sc xiii} 202.04~\AA~(bottom-row) have been shown in this figure. The locations of these two displayed Gaussian fitting samples in each line profile correspond to the pixel (0,36) and pixel (58,333). The standard 1-$\sigma$ error is also overplotted by red color in each sample fit.}
\label{fig:figure2}
\end{figure}
%++++++++++++++++++++++++++++++++++++++++++++++++++++++++++++++++++++++++++++++++++++++++++++++
%++++++++++++Table 1++++++++++++++++++++++++++++++++++++++++++++++++++++++++++++++++++++++++++++
\begin{table}
\caption{ This table shows  all the four lines, which are used in the present
analysis. Formation temperatures as well as the standard wavelengths  are also listed.  
}
\label{T-simple}
\begin{tabular}{ccclc}                               
  \hline                 
Sr. No. & Ion & Wavelength (\AA) & Log T (K)\\
  \hline
1 & Si VII  & 275.35  & 5.8\\
2 & Fe X    & 184.54  & 6.0\\
3 & Fe XII  & 195.12  & 6.1\\
4 & Fe XIII & 202.04  & 6.2\\
  \hline
\end{tabular}
\end{table}
%++++++++++++++++++++++++++++++++++++++++++++++++++++++++++++++++++++++++++++++++++++++++++++++++++++++++++++++++++++
 Si~{\sc vii} 275.35~\AA, Fe~{\sc x} 184.54~\AA\  and Fe~{\sc xiii} 202.04~\AA\ lines are comparatively clean lines and are not affected by blending issues \citep{Young2007}, while Fe~{\sc xii} 195.120~\AA\ is blended with Fe~{\sc xii} 195.180~\AA\ \citep{Young2009}. Although, \cite{Young2009} have shown that Fe~{\sc xii} 195.120~\AA\ is not affected appreciably by Fe~{\sc xii} 195.180~\AA\ line in the coronal holes, however, we have considered this blending and removed it in the present analysis.  Various instrumental effects related to the Hinode/EIS observations ({\it e.g.}, slit tilt, thermal variation, EIS CCD offset) must be addressed before the Gaussian fitting to derive the basis parameters from the observed line profiles. {\it Hinode}/EIS has single grating that disperse the light onto two CCDs and  spatial offset is present between these two CCDs of the {\it Hinode}/EIS. We have used standard {\it Hinode}/EIS routine, eis$\_$ccd$\_$offset.pro, to find out the pixel offset and direction of the offset ({\it i.e.}, up or down) in all lines relative to the He~{\sc ii} 256.32~\AA. After finding the amount and direction of the shift, we have shifted the observed spectra by that many pixels in the corresponding direction in all the lines. After that, we have selected a common region in all four lines for further analysis. To measure the thermal drift, we have used the same technique as \cite{Tian2010} have used in his analysis. After calculating slit tilt by using eis$\_$slit$\_$tilt.pro standard {\it Hinode}/EIS routine, we have built a wavelength correction offset array ({\it i.e.}, for wavelength correction) for all four lines. These wavelength offset arrays have the combined effect of slit tilt as well as thermal drift of respective spectral line. We have fitted a single Gaussian function on the three line profiles ({\it i.e.}, Si~{\sc vii}, Fe~{\sc x}  and Fe~{\sc xiii}). Double Gaussian has been fitted on the Fe~{\sc xii} line profile to separate the Fe~{\sc xii} 195.180~\AA\ component from Fe~{\sc xii}  195.12~\AA\ line profile. The combined effect of the slit tilt and orbital variation for any spectral line has been removed (wavelength vector corrected by applying wavelength correction offset array at each pixel of the observed region) before the Gaussian fitting in the chosen spectral line profile. 
%++++++++++++Figure 3+++++++++++++++++++++++++++++++++++++++++++++++++++++++++++++++++++++++++++++	
\begin{figure}    %%%%%%%%%%%%%%%%%% FIGURE 2
                                % includes the two top panels 
   \centerline{\hspace*{0.015\textwidth}
               \includegraphics[trim= 5.2cm 1.5cm 6.35cm 1.0cm, width=0.335\textwidth,clip=]{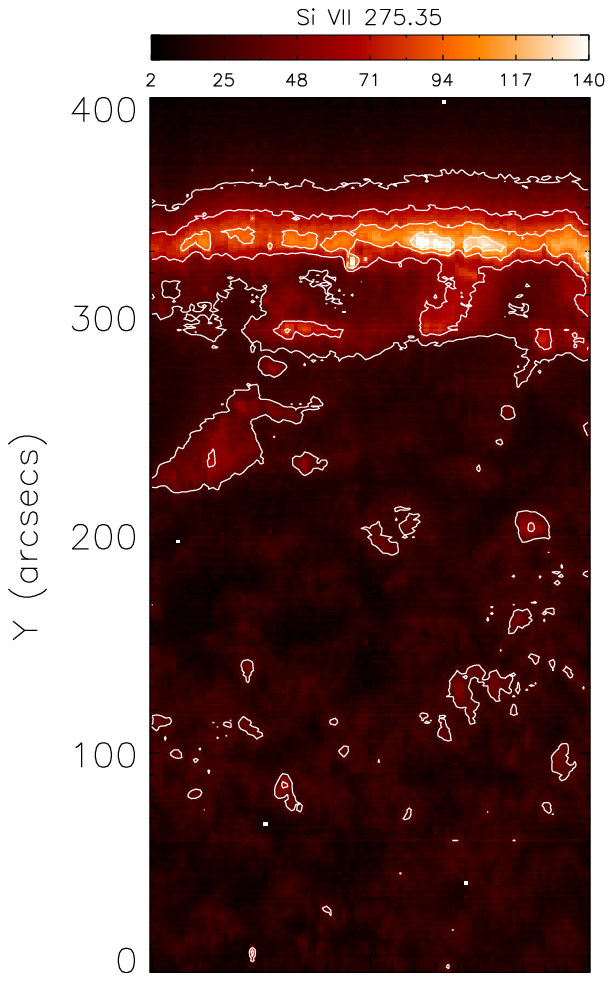}
               \hspace*{-0.09\textwidth}
               \includegraphics[trim= 5.2cm 1.5cm 6.35cm 1.0cm, width=0.335\textwidth,clip=]{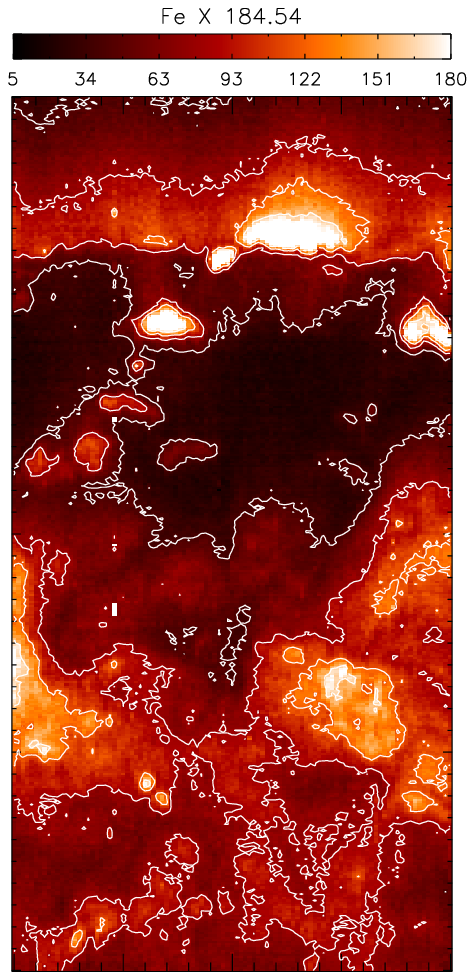} %This figures needs attention
               \hspace*{-0.09\textwidth}
               \includegraphics[trim= 5.2cm 1.5cm 6.35cm 1.0cm, width=0.335\textwidth,clip=]{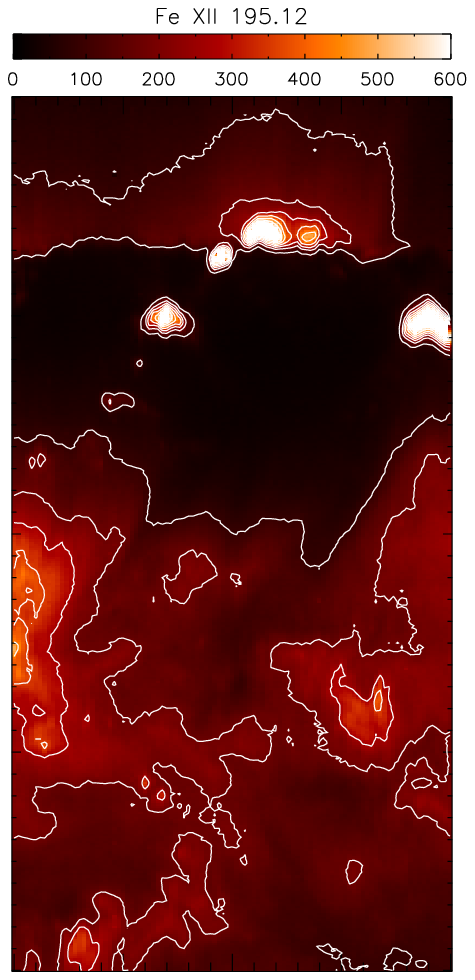}
               \hspace*{-0.09\textwidth}
               \includegraphics[trim= 5.2cm 1.5cm 6.35cm 1.0cm, width=0.335\textwidth,clip=]{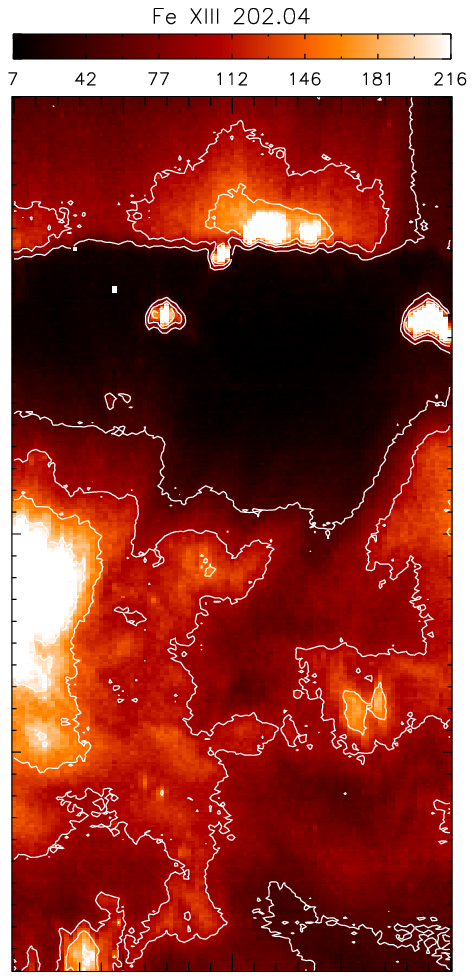}
              }
     \vspace{-0.35\textwidth}   % Shift close to the panel top 
     \centerline{\Large \bf     % Includes the labels (here needs the color 
                                %   package, see beginning of this file)
      %\hspace{0.0 \textwidth}  \color{white}{(a)}
      %\hspace{0.415\textwidth}  \color{white}{(b)}
         \hfill}
     \vspace{0.31\textwidth}    % Shift back to the panel bottom 
   \centerline{\hspace*{0.015\textwidth}
             \includegraphics[trim= 5.2cm 1.5cm 6.35cm 1.0cm, width=0.335\textwidth,clip=]{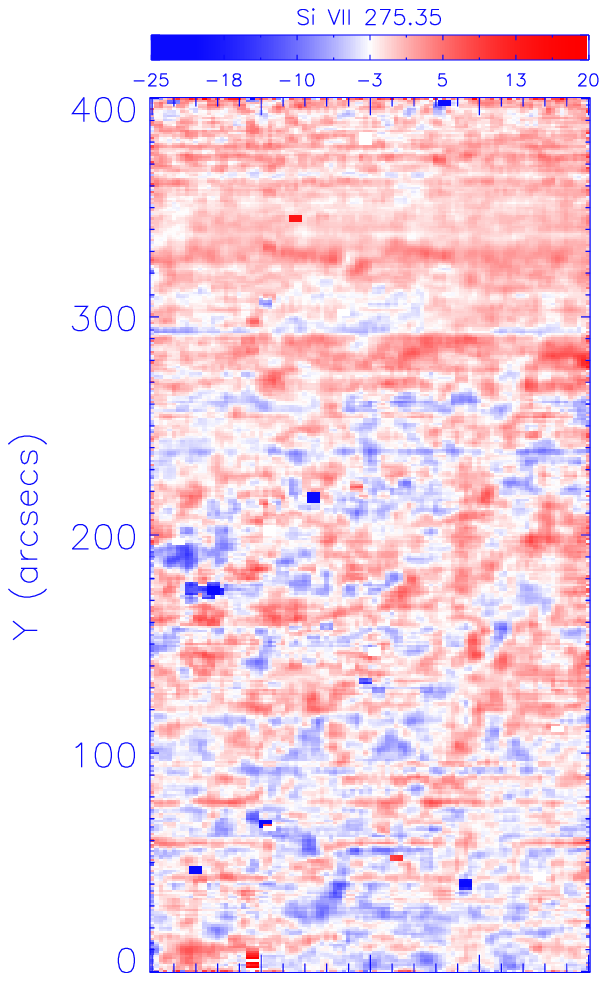}
             \hspace*{-0.09\textwidth}
             \includegraphics[trim= 5.2cm 1.5cm 6.35cm 1.0cm, width=0.335\textwidth,clip=]{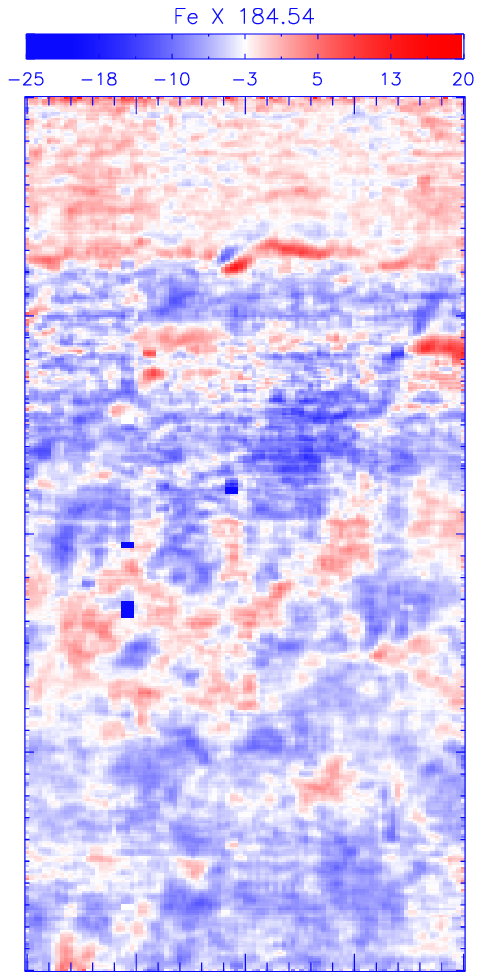}
             \hspace*{-0.09\textwidth}
             \includegraphics[trim= 5.2cm 1.5cm 6.35cm 1.0cm, width=0.335\textwidth,clip=]{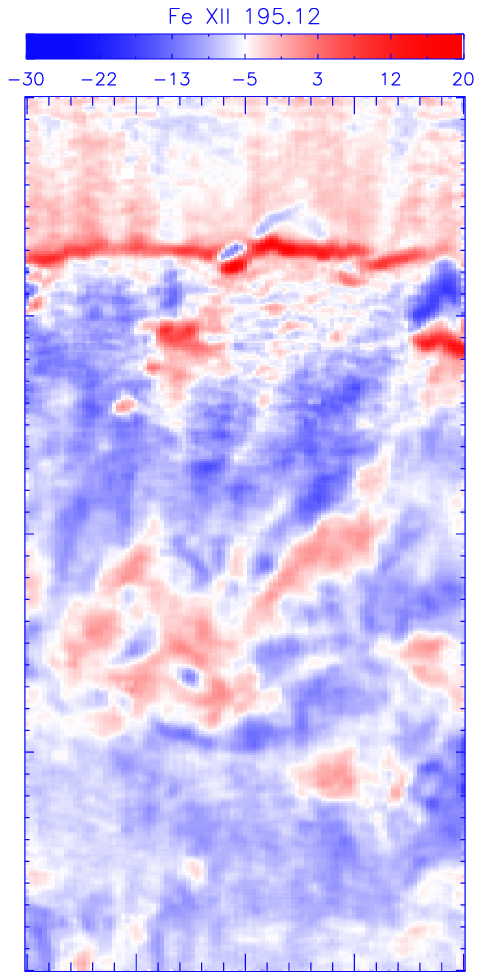}
             \hspace*{-0.09\textwidth}
             \includegraphics[trim= 5.2cm 1.5cm 6.35cm 1.0cm, width=0.335\textwidth,clip=]{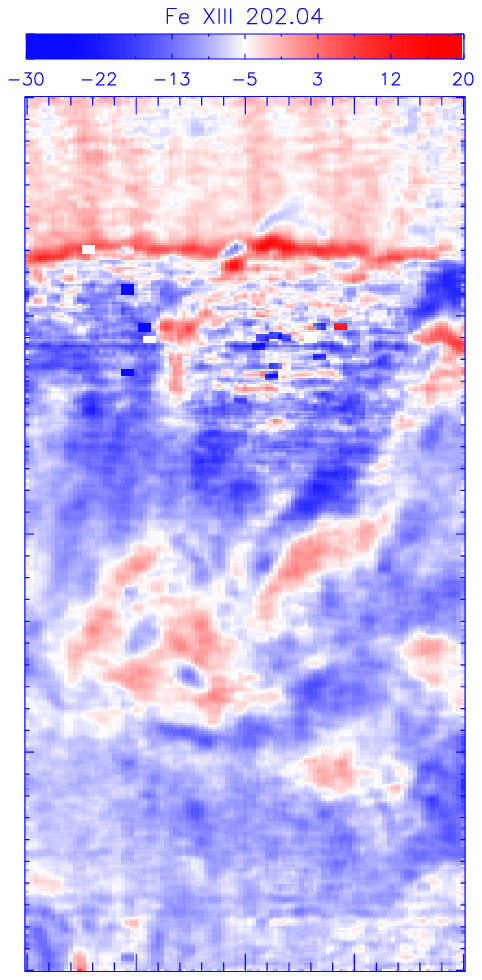}
              }
     \vspace{-0.35\textwidth}   % Shift close to the panel top 
     \centerline{\Large \bf     % Includes the labels (here needs the color package)
      %\hspace{0.0 \textwidth} \color{white}{(c)}
      %\hspace{0.415\textwidth}  \color{white}{(d)}
         \hfill}
     \vspace{0.31\textwidth}    % Shift back to the panel bottom 
\centerline{\hspace*{0.015\textwidth}
       %\hspace*{-0.11\textwidth}
             \includegraphics[trim= 5.2cm 0.8cm 6.35cm 1.0cm, width=0.335\textwidth,clip=]{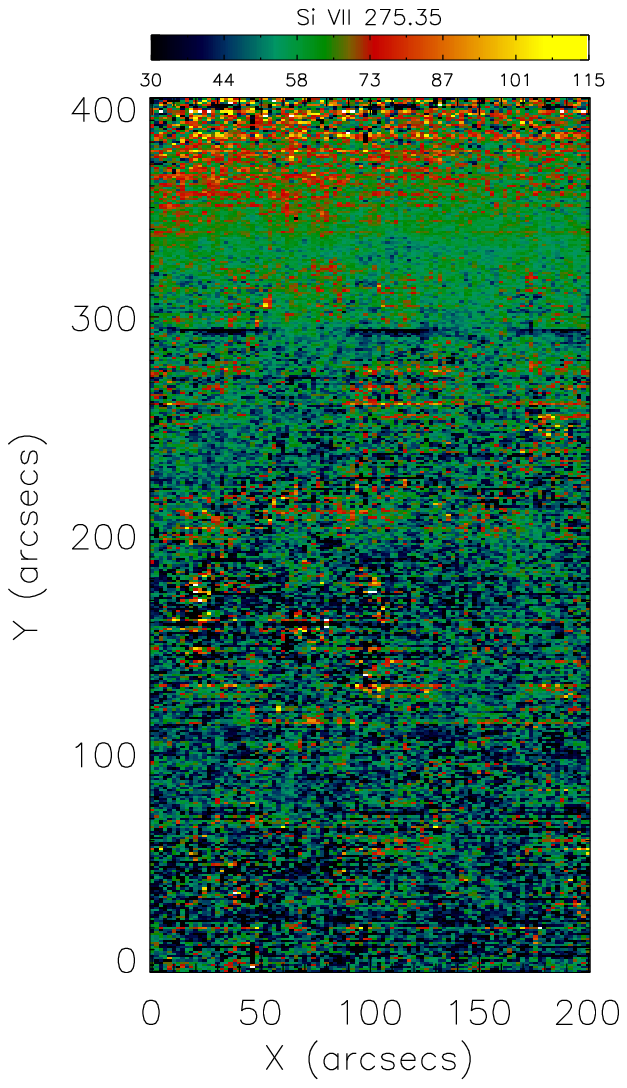}
              \hspace*{-0.09\textwidth}
              \includegraphics[trim= 5.2cm 0.8cm 6.35cm 1.0cm, width=0.335\textwidth,clip=]{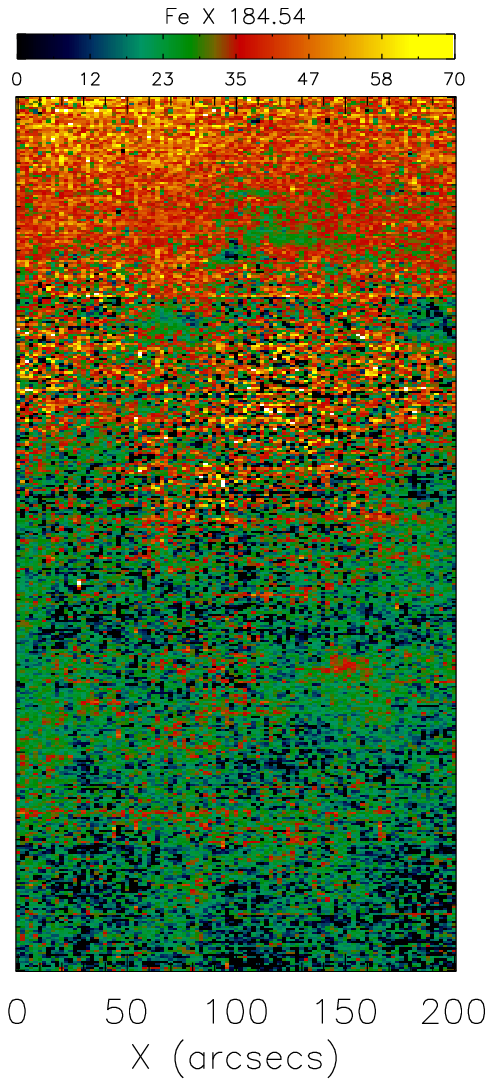}
              \hspace*{-0.09\textwidth}
             \includegraphics[trim= 5.2cm 0.8cm 6.35cm 1.0cm, width=0.335\textwidth,clip=]{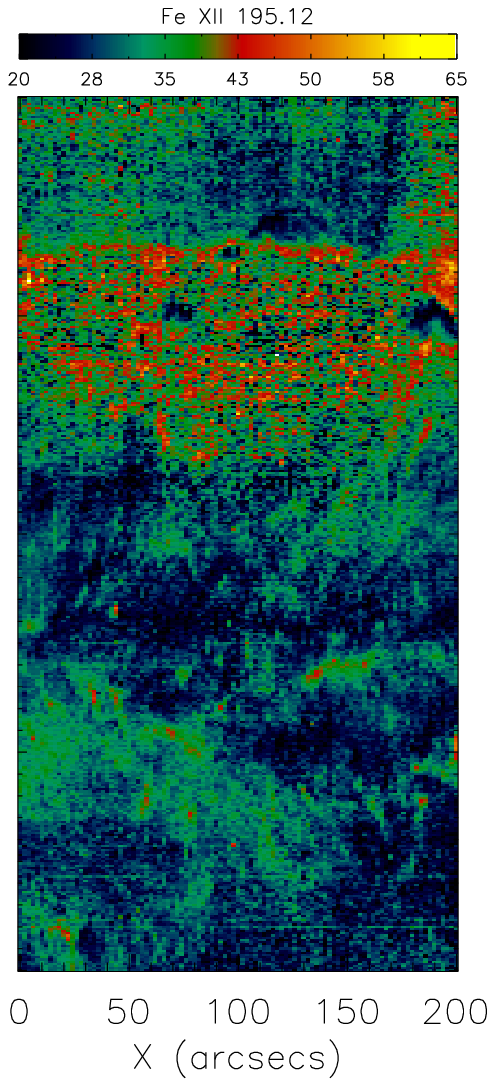}
              \hspace*{-0.09\textwidth}
              \includegraphics[trim= 5.2cm 0.8cm 6.35cm 1.0cm, width=0.335\textwidth,clip=]{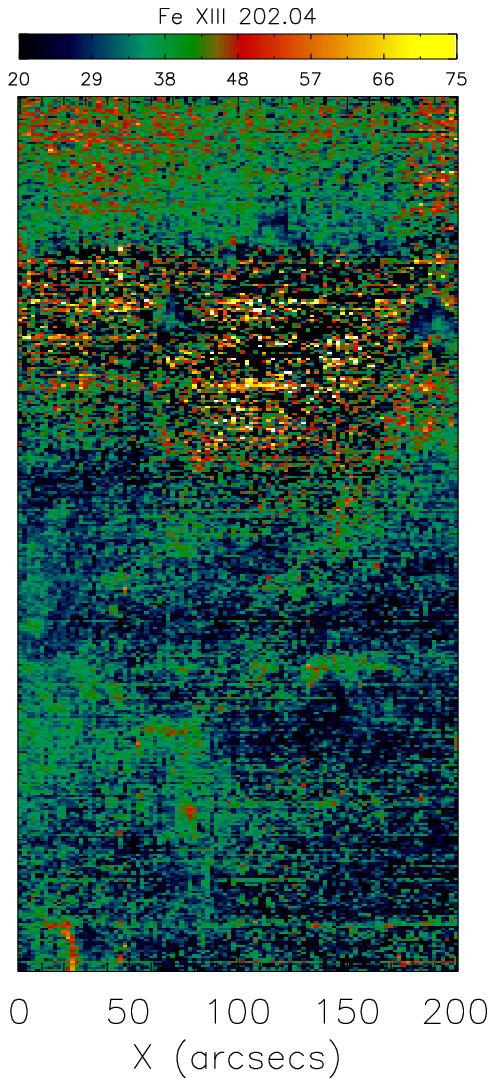}
}          
\caption{Intensity (top-row), Doppler velocity (middle-row) and FWHM (bottom-row) maps corresponding to the NPCH (as marked by a rectangular box in Figure~\ref{fig:fig1}) as observed by four lines, Si~{\sc vii} 275.5~\AA, Fe~{\sc x} 184.54~\AA, Fe~{\sc xii} 195.120~\AA\ and Fe~{\sc xiii} 202.04~\AA. Intensity contours are shown on the intensity maps.
        }
\label{fig:figure3}
\end{figure}
%+++++++++++++++++++++++++++++++++++++++++++++++++++++++++++++++++++++++++++++++++++++++++++++++++++++++++++++++

 For the double Gaussian fitting ({\it i.e.}, in case of Fe~{\sc xii} 195.120~\AA), we have used the procedure of \cite{Young2009}. Figure~\ref{fig:figure2} represents the two samples of Gaussian fit in each line profile at two different locations. From the Gaussian fitting, we have extracted the basic parameters (i.e., intensity, Gaussian sigma, line centeroid, etc.) at each location (pixels) of the commonly observed region. Rest wavelength measurement is difficult task as the absolute wavelength calibration is not available for {\it Hinode}/EIS. Unfortunately, any low temperature chromospheric line is not present in our wavelength window, therefore, in the present case we have used the Limb method \citep{PJ1999} for measuring the rest wavelengths. After measuring the rest wavelengths in each line profile, we have estimated the Doppler velocity at each location of the observed region. Similarly, we have estimated the FWHM from Gaussian sigma ({\it i.e.}, FWHM=2.3548$ \sigma$) over commonly observed region. The intensity, Doppler velocity and FWHM maps are shown in  Figure~\ref{fig:figure3} corresponding to all the four lines. Apart from the Doppler velocity, intensity and FWHM, electron density can also be estimated with the help of intensity ratio of two density sensitive lines and CHIANTI database \citep{Dere1997,Landi2006}. In the present analysis, we have used Fe~{\sc xii} 186.88~\AA\ and Fe~{\sc xii} 195.120~\AA\ for the density measurement, which represents the density at coronal heights. \cite{Young2009} have shown that both density sensitive Fe~{\sc xii} lines suffer from the blending problem ({\it i.e.}, Fe~{\sc xii} 186.88~\AA\ with S~{\sc xi} 186.839~\AA\ and Fe~{\sc  xii} 195.120~\AA\ with Fe~{\sc xii} 195.180~\AA). The intensity contribution of S~{\sc xi} 186.839~\AA\ in Fe~{\sc xii} 186.88~\AA\ intensity can be assessed from the S~{\sc xi} 188.617~\AA\ or S~{\sc xi} 191.266~\AA. On the basis of active region datasets, \cite{Young2009} have shown that S~{\sc xi} 186.839~\AA\ can contribute maximum 5\% intensity in Fe~{\sc xii} in the regime of moderate densities while S~{\sc xi} 186.839~\AA\ can contribute 2\% intensity in Fe~{\sc xii} 186.88~\AA\ in the regime of  more higher densities. Unfortunately, any one transition out of these two S~{\sc xi} transitions is not present in the present data-set, therefore, we have assumed that S~{\sc xi} 186.839~\AA\ contributes 5\% in the Fe~{\sc xii} 186.88~\AA\ intensity. Although, the assumption of the 5\% contribution of S~{\sc xi} 186.839~\AA\ in Fe~{\sc xii} 186.88~\AA\ is based on the active region data-sets but we have used this value (i.e., 5\% contribution) to remove the blending from Fe~{\sc xii} 186.88~\AA\ in the present NPCH observation because the intensity contribution of S~{\sc xi} 186.839~\AA\ is more effective towards the moderate densities in comparison to the very high densities \citep{Young2009}. To remove the blending from Fe~{\sc xii} 195.120~\AA~line, we have followed the same approach as described by \cite{Young2009}. After calculating the intensity ratio from these two density sensitive lines, we have derived the density over the observed common region ({\it cf.}, left panel of Figure~\ref{fig:figure7}). The density map (Figure~\ref{fig:figure7}) also represents the overplotted small boxes, which we have used for the present work and will be described in the upcoming section. We concentrated primarily on three regions in this work, which is shown by three dashed black boxes on the Doppler velocity map (Figure~\ref{fig:figure4}). These three regions represent quiet-Sun (QS; lower box), quiet-Sun with coronal hole boundary (QSCH; middle box) and coronal hole (CH; upper box) respectively. Measurement of the various plasma parameters have been evaluated for various localised regions that correspond to the three broadly classified physical regions as QS, QSCH and CH. 
\section{Comparison Between QS and CH Parameters} %%%%%%%%%%%%%%%%%%%%%%%%%%%%%%%%%%%%%%%%
      \label{S-QS-CH}      
%For the comparsion between Polar CH and adjacent QS, we have not selected the whole observed polar CH and QS, although, we have selected the some specific regions (i.e., some in excess FWHM and rest in average FWHM) in both observed QS and CH. As we know that the
Ion temperature and unresolved non-thermal motions  and/or presence of MHD waves are  responsible for the spectral broadening of optically thin coronal emission lines in the solar atmosphere. Generally, it is assumed that ions and electrons are in the thermal equilibrium, although the temperature of the ions and electrons may be different from each other, particularly at extended part of the corona. But for the inner corona, we have  assumed that ion temperature is identical to the electron temperature. FWHM can be expressed as follows,\\
\begin{equation}
FWHM = \bigg[W_{inst}+ 4\ln2 \bigg(\frac{\lambda}{c}\bigg)^2\bigg(\frac{2kT_i}{M_i}+\xi^2\bigg)\bigg]^{1/2},
\end{equation}
where, ${T_i}$, ${M_i}$, ${\xi}$ and ${W_{inst}}$ are the ion-temperature, ion-mass, non-thermal velocity, and instrumental-width respectively. Instrumental, thermal and non-thermal widths are the three components of the observed FWHM. Instrumental as well as thermal ({\it i.e.}, at a particular ion formation temperature) width components are constant, therefore, the variations in the FWHM arise due to the variation in the non-thermal width. The non-thermal width provides signature of various unresolved small-scale dynamics in the solar atmosphere, which do not normally recorded by imaging instruments.
%%%%%++++Figure 4++++++++++++++++++++++++++++++
\begin{figure*}[ht]
\centerline{\hspace*{0.15\textwidth}
               \includegraphics[height=0.60\textheight, width=1.6\textwidth,clip=]{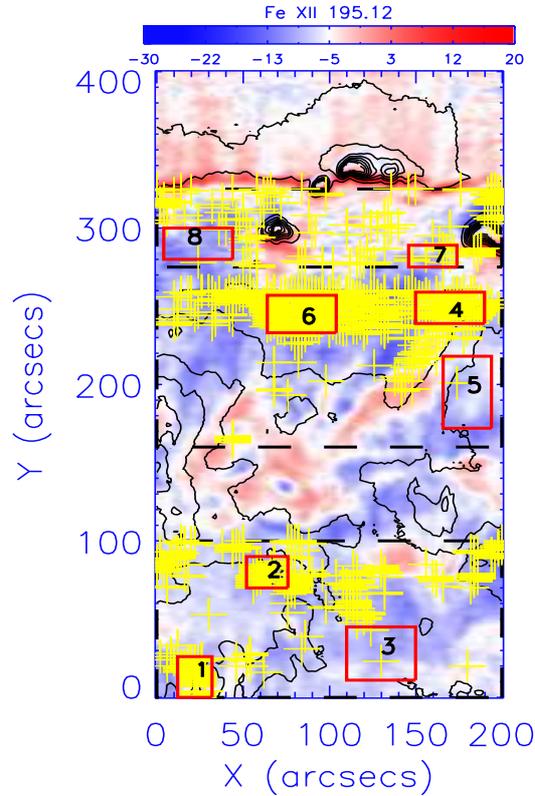}
}
\caption{The image shows Doppler velocity map of the observed region, which is overlaid by intensity contours as well as excess width locations (marked by yellow plus symbols). All selected regions have been over-plotted on the Doppler velocity map by red rectangular boxes.}
\label{fig:figure4}
\end{figure*}
%***************************************************
Figure~\ref{fig:figure4} shows the Doppler velocity map in Fe~{\sc xii} 195.120~\AA\ line, which is overlaid with intensity contours.  The disk part of the observed FOV is divided into three parts for  further analysis, which is also marked on the Doppler velocity map and demarcated by black dashed lines (Figure~\ref{fig:figure4}). From bottom to top, these three regions represent quiet-Sun (QS), quiet-Sun with Coronal Hole boundary (QSCH) and Coronal Hole (CH) respectively. The locations where we observe excess widths compared to its average values in the neighbouring regions may allow us to locate these locations where various transients/explosive events are occurring. Furthermore, we have searched the excess width locations in all three regions by setting  different threshold level in each box. Widths above the threshold levels have been considered as the excess widths. The average widths as well as the threshold levels in  all the three regions ({\it i.e.}, QS, QSCH and CH) have been listed in Table~\ref{T-excess width}. After finding the excess width locations, we have marked these locations over the Doppler velocity map by yellow plus signs ({\it cf.}, Figure~\ref{fig:figure4}).   
%++++++++++++Table 2++++++++++++++++++++++++++++++++++++++++++++++++++++++++++++++++++++++++++++
\begin{table}
\caption{This table shows the average observed widths of all three region (i.e., QS, QSCH and CH). Similarly, the adopted threshold level of widths  in each box to locate the excess widths in these regions have also been shown in this table.   
}
\label{T-excess width}
\begin{tabular}{cccc}                                
\hline                 
Sr. No. & Big Black Dashed Box & Average Width (m\AA) & Threshold Width (m\AA)\\
\hline
1 & QS & 33.65  & 1.20$\times$Average Width\\
2 & QSCH & 33.47  & 1.30$\times$Average Width\\
3 & CH   & 39.42  & 1.35$\times$ Average Width\\
\hline
\end{tabular}
\end{table}
%+++++++++++++++++++++++++++++++++++++++++++++++++++++++++++++++++++++++++++++++++++++++
Finally, we have selected eight different boxes of different sizes in these three regions. Some boxes, out of these eight boxes, correspond to the excess width regions while rest of the boxes are located in the average width regions. All the eight selected boxes are overplotted on the same Doppler velocity map ({\it cf.}, Figure~\ref{fig:figure4}) by dark red color and are numbered. The  description of all the selected regions ({\it i.e.}, three big boxes as well as eight small boxes) is given in Table~\ref{table:boxes}. We have chosen these small eight regions to compare various parameters corresponding to the excess width locations with the locations having average width below the threshold values. On the basis of this comparison, we have performed a comparison between QS and CH as well as tried to locate the coronal funnels in QS, QSCH and CH.
%+++++++++++++Table 3++++++++++++++++++++++++++++++++++++++++++++++++++++++++
\begin{table}[ht]
\caption{The table gives the details about the all selected boxes, which is used in the present analysis.}
\begin{tabular}{|l|l|l|}
\hline
\multicolumn{3}{ |c| }{Description of different selected regions} \\
\hline
Big Black Dashed Boxes & Small Boxes & Description \\ \hline
\multirow{3}{*}{QS (Lower Box)} & Box 1 & Excess Width \\
 & Box 2 & Excess Width\\
 & Box 3 & Average Width \\ \hline
\multirow{3}{*}{QSCH (Middle Box)} & Box4 & Excess Width \\
 & Box 5 & Average Width \\
 & Box 6 & Excess Width \\ \hline
\multirow{2}{*}{Coronal Hole (Upper Box)} & Box 7 & Excess Width\\
 & Box 8 & Average Width \\ \hline
\end{tabular}
\label{table:boxes}
\end{table}
%+++++++++++++++++++++++++++++++++++++++++++++++++++++++++++++++++++++++++++++++++++++++
\subsection{Doppler Velocity Variations} %%%%%%%%%%%%%%
\begin{figure*}[ht]
\centerline{\hspace*{0.01\textwidth}
            \includegraphics[trim= 0.5cm 0.25cm 0.30cm 0.3cm, height=0.2\textheight, width=0.5\textwidth,clip=]{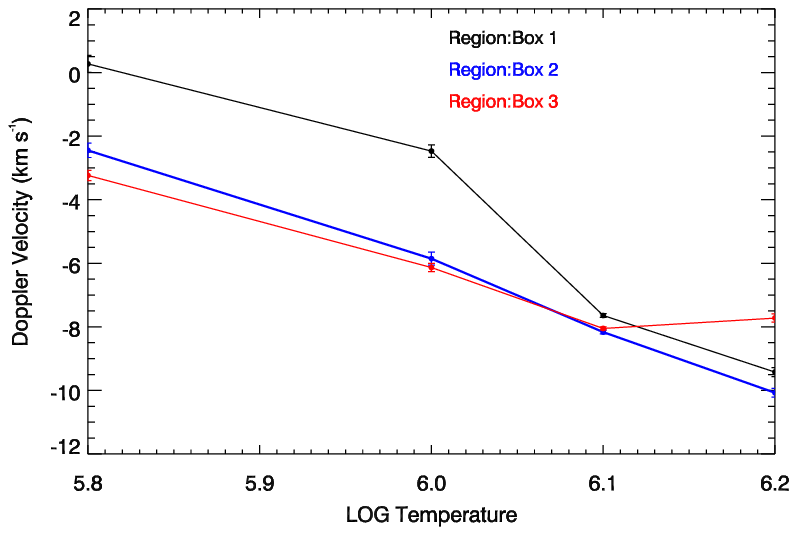}
            \includegraphics[trim= 0.5cm 0.25cm 0.30cm 0.3cm, height=0.2\textheight, width=0.5\textwidth,clip=]{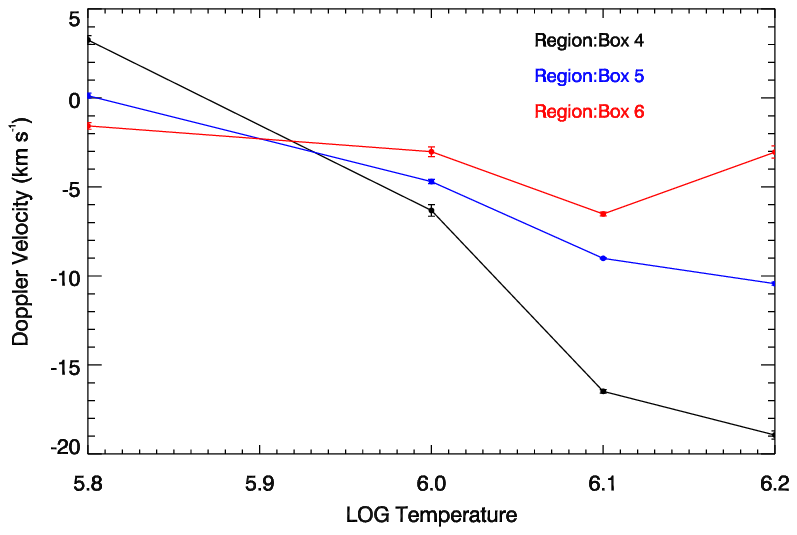}
}
\centerline{\hspace*{0.01\textwidth}
           \includegraphics[trim= 0.5cm 0.25cm 0.30cm 0.3cm, height=0.2\textheight, width=0.5\textwidth,clip=]{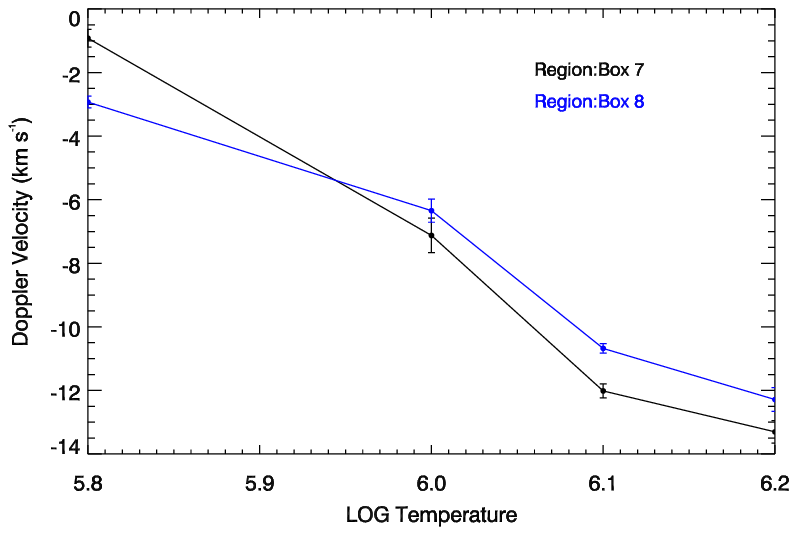}
           \includegraphics[trim= 0.5cm 0.25cm 0.30cm 0.3cm, height=0.2\textheight, width=0.5\textwidth,clip=]{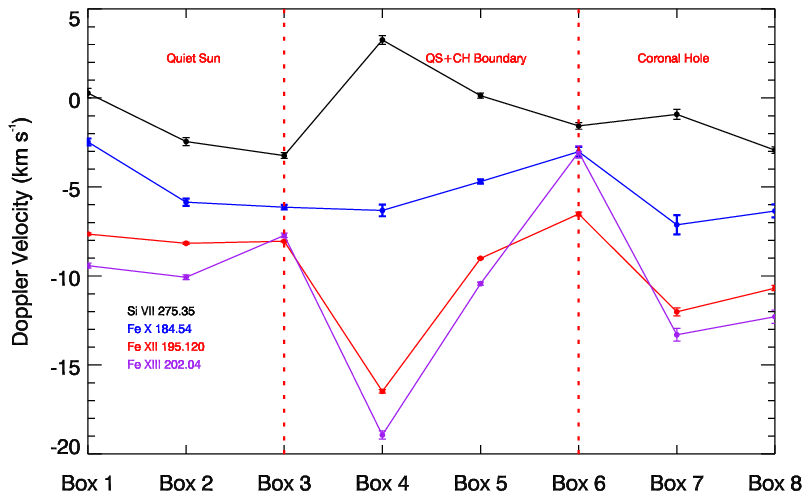}
}
\caption{Doppler velocity variations with temperature corresponding to different regions QS (top-left), QSCH (top-right) and CH (bottom-left). 
The overall trend is shown in bottom-right panel. }
\label{fig:figure5}
\end{figure*}

  \label{S-Doppler Velocity}
Investigation of the variation of Doppler velocity with temperature reveals the nature of plasma flow ({\it i.e.}, up-flows/down-flows) in the solar atmosphere. The lines, which we have used in the present study, cover the  TR (Si~{\sc vii}  275.35~\AA, Log T$_{e}$ = 5.8 K) up to the solar corona (Fe~{\sc xiii} 202.04~\AA, Log T$_{e}$ = 6.2 K), corresponding to the QS (upper left, Figure~\ref{fig:figure5}), QSCH (upper right, Figure~\ref{fig:figure5}) and CH (bottom left, Figure~\ref{fig:figure5}). The overall trend is shown in the right panel of Figure~\ref{fig:figure5}. 
%++++++++++++++++++++++Figure5++++++++++++++++++++++++++++++++++++++++++++++++++++++
%++++++++++++++++++++++++++++++++++++++++++++++++++++++++++++++++++++++++++++++++++++
The Doppler velocities corresponding to the observed excess width regions of QS ({\it i.e.}, Box 1 and 2) are $\sim$ 0.26$\pm$0.27 and $\sim$ -2.44$\pm$0.20 km s$^{-1}$ at Si~{\sc vii} formation temperature, therefore, these excess width regions show marginal up-flows in upper TR. The plasma up-flows in these excess width regions  increases with temperature up to the corona, which have maximum Doppler velocity of $\sim$ -9.42$\pm$0.14 km s$^{-1}$ and $\sim$ -10.07$\pm$0.14 km s$^{-1}$ in both excess width regions ({\it i.e.}, Box 1 and 2) at the temperature Log T$_{e}$ = 6.2 K ({\it cf.}, Figure~\ref{fig:figure5}). The Doppler velocity of Box 3, which is located in the average width region of QS, is $\sim$ -3.24$\pm$0.16 km s$^{-1}$ at Si~{\sc vii} ion formation temperature, which increases further in the solar atmosphere up to Fe~{\sc xii} formation temperature with the blue-shift value $\sim$ -8.05$\pm$0.05 km s$^{-1}$. However, at the Fe~{\sc xiii} 202.04~\AA\ ion formation temperature, this region shows slight decrease in the blue-shift ({\it i.e.}, -7.7$\pm$0.13 km s$^{-1}$). 
The reduction in blue-shift at Fe~{\sc xiii} 202.04~\AA\ from Fe~{\sc xii} 195.120~\AA\ is not very high in this average width region of QS, although, blue-shift may decrease further towards higher heights in the solar atmosphere. Therefore, we can say that the Doppler velocity in the average width region of QS (Box 3) may have a different nature from the Doppler velocity nature of excess width regions of the QS (Box 1 and 2). It should be noted that we are trying to identify these excess width regions as the footprints of coronal funnels, which should show predominance of either wave activity and/or flows.\\
Similar to the QS region, in the QSCH region (upper-right snapshot, Figure~\ref{fig:figure5}), we have also selected three boxes for similar purpose. Boxes 4 and 6 are located in the excess width regions while Box 5 corresponds to the average width region ({\it cf.}, Table~\ref{table:boxes}). Box 4, which is located at the boundary of CH, shows down-flows ({\it i.e.}, red-shifts) at TR and as the temperature increases the red-shift inverts into blue-shift with the maximum value of $\sim$ -18.93$\pm$0.23 km s$^{-1}$ at Log T$_{e}$ = 6.2 K. On contrary, the Doppler velocity of Box 6, which is also an excess width region, does not change very much from Si~{\sc vii} 275.35~\AA\ ion (-1.57$\pm$0.18 km s$^{-1}$) to Fe~{\sc xiii} 202.04~\AA\ (-3.04$\pm$0.34 km s$^{-1}$). Therefore, the Doppler velocity is almost constant in this box from TR up to the solar corona. The Doppler velocity of Box 5, which is an average width QS region near CH boundary, is weakly red-shifted (0.13$\pm$0.14 km s$^{-1}$) at Log T$_{e}$ = 5.8 K. After that the up-flow speed increases with temperature with a maximum value of -10.43$\pm$0.09 km s$^{-1}$ at Log T$_{e}$ = 6.2 K. Therefore, it is very difficult to make distinction between the excess and average width regions on the basis of the net Doppler velocities in the QSCH regions. \\
%It seems that the average width box (i.e., box 5) has inversion point from red-shift to blue-shift at Log T$_{e}$ = 5.8 K, although, excess width box 4 has inversion point from red-shift to blue-shift at Log T$_{e}$ \textgreater\ 5.8 K while the excess width box 6 is already blue-shifted at Si~{\sc vii} 275.35~\AA\ ion formation temparature.\\
 Excess width (Box 7) and average width (Box 8) regions of CH ({\it cf}., Table~\ref{table:boxes}) have the Doppler velocities of -0.92$\pm$0.28 km s$^{-1}$ and -2.93$\pm$0.19 km s$^{-1}$ in Si~{\sc vii} at Log$_{e}$ = 5.8, respectively. The blue-shifts increase with temperature in both the regions and the maximum values of blue shifts are -13.30$\pm$0.35 km s$^{-1}$ and -12.29$\pm$0.38 km s$^{-1}$ in these excess as well as average width boxes at Log T$_{e}$ = 6.2 K. %As the Si~{\sc vii} lines is slightly blue-shifted in both boxes of CH, therefore, plasma flow inversion takes place below Si~{\sc vii} ion formation temperature. 
Again, it is very difficult to distinguish excess width location (Box 7) from average width location (Box 8) because both types of regions show similar pattern of the Doppler velocity. One can conjecture that it may be that corresponding to CHs the funnels have expanded so much at these heights, therefore, one can not distinguish a funnel and inter funnel regions. \\
%we have selected two boxes (cf., Bottom left, Fig.~\ref{fig:figure5}) and out of these two boxes one box (i.e., box 7) is located in the excess width while another box (i.e., box 8) is located in the average width region. The excess as well as average width boxes. Plasma up-flows (i.e., Doppler velocity inversion from red-shift to blue-shift) has been started below the Si~{\sc vii} 275.35~\AA\ ion formation temparature in the excess as well as avearge width regions of CH as both regions show blue-shifts at the Si~{\sc vii} 275.35~\AA\ ion formation temparature.
The overall comparison of Doppler velocity of each emission line in all the eight boxes (Bottom right panel of Figure~\ref{fig:figure5}) shows that  coolest line({\it i.e.}, Si~{\sc vii} 275.35~\AA, Log T$_{e}$=5.8 K) is slightly blue-shifted in the QS, QSCH and CH. Although, the Box 5, which is located at the boundary of CH and QS (see Figure~\ref{fig:figure4}), shows larger red-shift at this temperature. The higher temperature lines ({\it i.e.}, Fe~{\sc x} 184.54~\AA, Fe~{\sc xii} 195.120~\AA\ and Fe~{\sc xiii} 202.04~\AA) move towards the higher blue-shifts as the temperature rises. Therefore, the cool line is dominated by down-flows, while the hot lines are dominated by up-flows as already been reported in the literatures (cf., Figure 6). 
%and the most hot line (i.e., Fe~{\sc xiii} 202.04~\AA\, Log T$_{e}$ = 6.2 K) in the present analysis shows that all the selected boxes of CH region has slightly higher blue-shift in comparsion to the blue-shifts of the all selected boxes of QS. 
%\subsubsection{Average Doppler-shift from Chromosphere to Corona in QS}
%\label{S-T_DV}

\begin{figure}[ht]
\centerline{\includegraphics[height=0.4\textheight, width=1.0\textwidth,clip=]{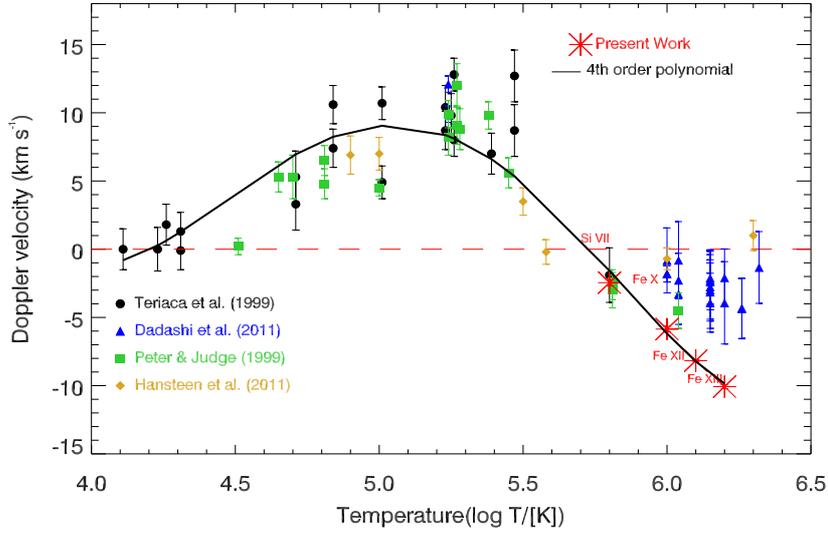}
              }
\caption{Average Doppler shift in the QS at disk center corresponding to various TR and CH ions as measured by SUMER and EIS spectra and as reported by earlier works as labeled. Our estimated Doppler velocity results of QS are marked by red-star symbols. Positive values indicate red-shifts (down-flows), while negative values indicate blue-shifts (up-flows). The solid line represents a fourth order polynomial fit to the data.  
                  }
\label{fig:figure6}
\end{figure}

 The average Doppler shift of spectral lines formed from chromosphere to the corona revel important information on the mass and energy balance of the solar atmosphere. We compare our results with previously published net Doppler shift values (cf., Figure~\ref{fig:figure6}), corresponding to the quiet-Sun for various TR and coronal lines as recorded by SUMER and EIS spectra. Positive values indicate red-shifts (down-flows), while the negative values indicate blue-shifts (up-flows). 
%%++++++++++++Figure 6+++++++++++++++++++++++++++++++++++++++++++++++++  
%%%*****************************************************************************
The QS Doppler velocities, which we have obtained in the present work by using Hinode/EIS spectrum, are marked by red-star symbols. We should also point out that absolute values of the Doppler velocities depend on the rest wavelength and the measurement of the rest wavelength is very crucial. As we have described earlier in Section~\ref{S-data-analysis} that we have used the limb method \citep{PJ1999, Dam1999} to calculate the accurate rest wavelengths of the used EIS lines. In the limb method, it is assumed that the average Doppler velocity is almost zero around the solar limb ({\it i.e.}, motions along the line of sight cancel out on average in an optically thin plasma). For the 4$^{th}$ order polynomial fit (solid black line) form chromosphere to the corona, we have include results from \citealt{Teriaca1999}. 

\subsection{Electron Density Variations}
\begin{figure}[ht]
\centerline{
\hspace{-0.3\textwidth}
\includegraphics[height=0.45\textheight, width=1.0\textwidth, clip=true]{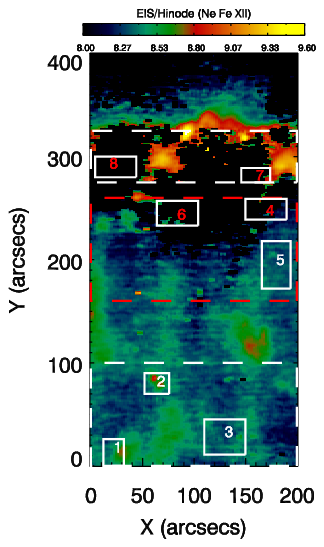}
\hspace{-0.45\textwidth}
\includegraphics[height=0.3\textheight, width=0.7\textwidth,clip=]{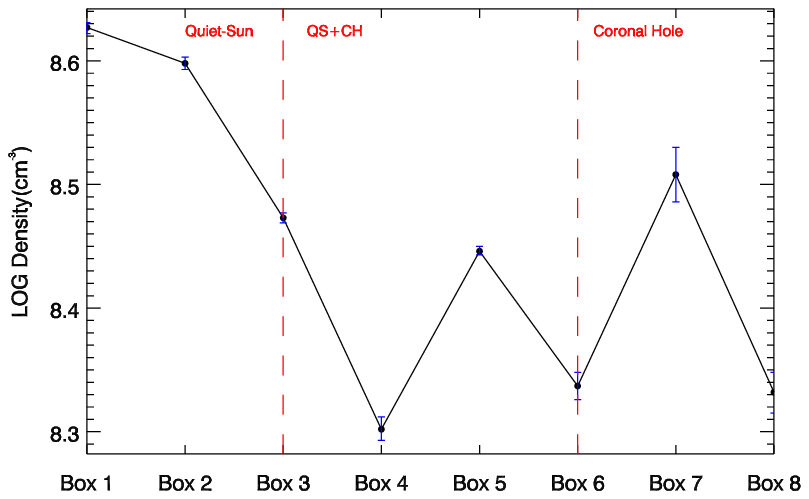}
}
\caption{The electron density map of the observed (left panel). The small selected regions are overplotted on the electron density map by white rectangular boxes. The averaged electron densities of all boxes are shown in the right panel.    
}
\label{fig:figure7}
\end{figure}

 The best available density sensitive Fe~{\sc xii} line pair has been used for the density measurement (Section~\ref{S-data-analysis}). The left panel of Figure~\ref{fig:figure7} represents the density map of the observed region while the right panel shows averaged density corresponding to the three QS boxes, three QSCH boxes and two CH boxes (for box description; see Section~\ref{S-QS-CH}). The density decreases progressively from QS to CH ({\it cf.}, right panel; Figure~\ref{fig:figure7}), which is expected. In the QS region, the electron densities of the excess width boxes (i.e., Box 1 and 2) are Log N$_{e}$ = 8.63 cm$^{-3}$ and Log N$_{e}$ = 8.60 cm$^{-3}$ respectively, while the electron density of the average width box ({\it i.e.}, Box 3) is Log N$_{e}$ = 8.47 cm$^{-3}$. Therefore, the excess width boxes have the higher electron densities in comparison to the density of average width box within QS.\\
%++++++++++++++++Figure 7+++++++++++++++++++++++++++++++++++++++++++++++++++++++++++
%++++++++++++++++++++++++++++++++++++++++++++++++++++++++++++++++++++++++++++++++++

 In QSCH region ({\it cf.}, QS+CH dashed box of right panel of Figure~\ref{fig:figure7}), the electron density of the excess width boxes (i.e., Box 4 and 6) are Log N$_{e}$ = 8.30 cm$^{-3}$ and Log N$_{e}$ = 8.33 cm$^{-3}$ respectively, while the electron density of average width box ({\it i.e.}, Box 5) is Log N$_{e}$ = 8.45 cm$^{-3}$. The electron density of average width box is high in comparison to the excess width boxes because the average width box is located in the quiet-Sun while the excess width boxes are located in the Qs-CH and CH regions. Coronal hole boundaries are the most probable locations of various explosive events \citep{Mad2004,Mad2009,Mad2012}, therefore, we expect that the excess width as well as high density locations lie along the coronal hole boundaries. Excess widths are present along the coronal hole boundary ({\it cf.}, yellow plus signs in Figure\ref{fig:figure4}). Although, the density of Box 4 that covers the coronal hole boundary along with large portion of the coronal hole, does not represent high density compared to the density of the box completely lying within coronal hole ({\it i.e.}, Box 6). Presence of the higher density (left panel of Figure~7) as well as maximum number of excess width pixels within Box 6 are to be noted.
In CH, the density of the excess (Box 7) and average (Box 8) width locations are Log N$_{e}$ = 8.50 cm$^{-3}$ and Log N$_{e}$ = 8.34 cm$^{-3}$ respectively. The higher electron density corresponding to  the excess width box (Box 7) may be influenced by the presence of a bright point (see left panel of Figure~\ref{fig:figure7}). Therefore, if we neglect the presence of the high density regions from the Box 5 and Box 7, then we can assume that the electron density of all excess  and average width boxes located in coronal hole ({\it i.e.}, Box 4, 6, 7 and 8) are almost same. On the basis of the electron density, it is not possible to make distinction between excess and average width regions of CH and QSCH.
%there is no discrimination between the excess and average width regions of the CH as we have found the discrimanation between the excess and average width regions in QS.
 
\section{Temperature Dependent Behaviour of Non-Thermal Velocity in QS}
\label{S-T-DV_NTV}
%Non-thermal velocities, which have been consistently observed in solar EUV spectral observations by using various space-borne observations, have been theorized to result from many plausible scenarios including wave motions, turbulence, or magnetic reconnection. 
 Non-thermal velocities can provide information about the unresolved motions/MHD waves in the solar atmosphere. Instrumental, thermal and non-thermal widths are the three component of the observed line width. Therefore, to get non-thermal width, the removal of the instrumental and thermal widths from the line width has been performed by using following formula,
\begin{equation} 
W_{nt} = [W_{obs}^{2}-W_{inst}^{2}-W_{th}^{2}]^{1/2},
\end{equation}
where, W$_{nt}$, W$_{obs}$, W$_{inst}$ and W$_{nt}$ are non-thermal, observed, instrumental and thermal widths respectively. The on-average instrumental width (i.e., W$_{inst}$) is $\sim$ 66 m\AA\ while the thermal width can be calculated using the relation,
\begin{equation}
 W_{th} = 4\ln2 \bigg(\frac{\lambda}{c} \bigg)^2 \bigg[\frac{2kT_i}{M_i}\bigg].
\end{equation}
%++++++++++++++++++++Figure 8++++++++++++++++++++++++++++++++++++++
%\begin{figure}
%\centerline{
%\includegraphics[trim= 0.5cm 0.1cm 0.2cm 0.4cm, height=0.28\textheight, width=0.65\textwidth,clip=]{fig8a.eps}
%\includegraphics[trim= 0.5cm 0.1cm 0.2cm 0.4cm, height=0.28\textheight, width=0.65\textwidth,clip=]{fig8b.eps}
%}
%\centerline{
%\includegraphics[trim= 0.5cm 0.1cm 0.2cm 0.0cm, height=0.28\textheight, width=0.65\textwidth,clip=]{fig8c.eps}
%\includegraphics[trim= 0.5cm 0.1cm 0.1cm 0.0cm, height=0.28\textheight, width=0.65\textwidth,clip=]{fig8d.eps}
%}
%\caption{
%}
%\label{fig:figure8}
%\end{figure}
%All small selected boxes, which are related to the QS, QSCH and CH, show a decrease in the non-thermal width from upper TR upto the solar corona, although, the non-thermal width in the QSCH and CH regions are higher in comparsion to the QS (cf., bottom left:Fig.~\ref{fig:figure8}). For the meaursement of non-thermal velocities, we followed a \\
%These non-thermal velocities (i.e., converted from non-thermal widths) provides clues about nature of plasma flow/waves in the solar atmosphere. QS region non-thermal widths and 
After estimating non-thermal widths, we have calculated the non-thermal velocities of QS and CH for Si~{\sc vii} 275.35~\AA, Fe~{\sc xii} 195.120~\AA\ and Fe~{\sc xiii} 202.04~\AA\ lines. The non-thermal velocity is calculated from this equation, 
\begin{equation}
\xi = \bigg(\frac{W_{nt}^{2}c^2}{4\ln2\lambda^2} \bigg)^{1/2} 
\end{equation}
%*****************QS Non-Thermal Velocities Table************************
\begin{table}[ht]
\caption{Quiet Sun Non-thermal Velocity $\xi$ (km s$^{-1}$)}
\begin{tabular}{c c c c}
\hline
Box number & Si VII & Fe XII & Fe XIII \\
\hline
1 &  23.64$\pm$0.35 & 28.76$\pm$0.07 & 22.23$\pm$0.02\\
2 &  24.34$\pm$0.28 & 31.48$\pm$0.07 & 20.09$\pm$0.17\\
3 &  27.64$\pm$0.21 & 31.45$\pm$0.05 & 20.92$\pm$0.16\\
\hline
\end{tabular}
\label{table:QS_NTV}
\end{table}
%*******************************************************************
QS region non-thermal velocities of Si~{\sc vii}, Fe~{\sc xii} and Fe~{\sc xiii} lines have been listed in Table~\ref{table:QS_NTV} while the CH non-thermal velocities  have been listed in the Table~\ref{table:CH_NTV}. The non-thermal velocities follow the same pattern in all selected boxes ({\it i.e.}, located in QS and CH regions) within the narrow temperature range (from Si~{\sc vii} up to Fe~{\sc xiii} ion temperature) in the solar atmosphere.
%******************CH Non-thermal Velicitis***************************************
\begin{table}[ht]
\caption{Coronal Hole Non-thermal Velocity $\xi$ (km s$^{-1}$)}
\begin{tabular}{c c c c}
\hline
Box number & Si VII & Fe XII & Fe XIII \\
\hline
7 &  33.28$\pm$0.35 & 37.83$\pm$0.26 & 24.26$\pm$0.44\\
8 &  31.13$\pm$0.24 & 37.76$\pm$0.18 & 26.72$\pm$0.49\\
\hline
\end{tabular}
\label{table:CH_NTV}
\end{table}
%***************************************************************************
\begin{figure}[ht]
\centerline{
\includegraphics[height=0.40\textheight, width=1.0\textwidth,clip=]{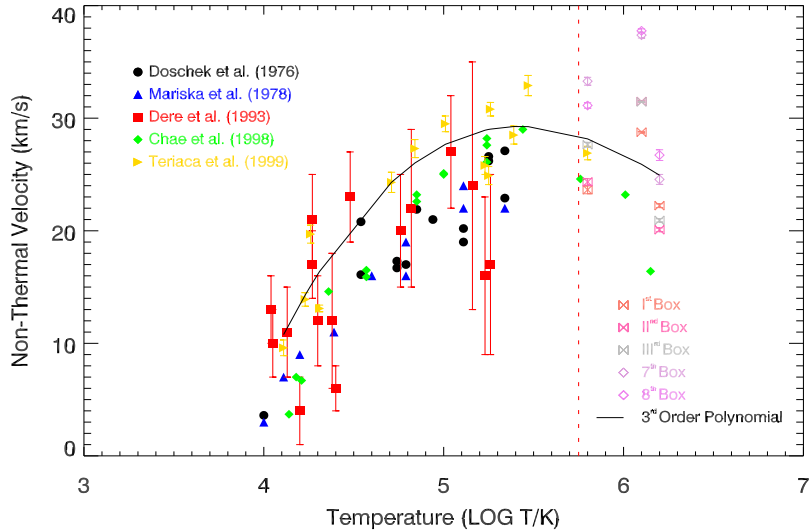}
}
\caption{ Variation of the non-thermal velocities with temperature corresponding to the QS. To cover a wide range of temperatures, we have included previous results as labeled. Using non-thermal velocities reported by \citealt{Teriaca1999} and derived from Box 3, we have fitted a 3$^{rd}$ order polynomial as shown by solid black line. Our estimated results of QS and CH boxes are plotted to the right side of the red vertical dashed line.   
}
\label{fig:figure8}
\end{figure}
%*************************************************************************************
The non-thermal velocity increase from Si~{\sc vii} up to the Fe~{\sc xii} and after that it decreases towards Fe~{\sc xiii} ion in the all selected boxes. We have included previous results as reported in  \cite{Doschek1976, Mariska1978, Dere1993, Chae1998}. This plot covers a large portion of the solar atmosphere and allows us to compare our results with previous numbers. It also provides a new unified physical scenario in the solar atmosphere which shows how the non-thermal velocity behave between 4.0 K $\leq$ Log T$_{e}$ $\geq$ 6.2 K, {\it i.e.}, covering from chromosphere, TR, corona above QS. The behaviour of non-thermal velocity with temperature in all selected QS ({\it i.e.}, 1$^{st}$, 2$^{nd}$ and 3$^{rd}$ ) boxes along with  previous results has been shown in Figure~\ref{fig:figure8}. We have also plotted the non-thermal velocities of CH boxes ({\it i.e.}, 7$^{th}$ and 8$^{th}$ boxes) in the same plot.
%as well as CH (i.e., 7$^{th}$ and 8$^{th}$ boxes) boxes along with various previous results has been shown in figure~\ref{fig:figure8}. The previous results of the non-thermal velocities are related to the quiet-sun, therefore, these results should be used only with QS region non-thermal velocities. Although, we have used the CH regions non-thermal velocities along with these previous QS non-thermal velocities, however, we are intersted to see that how these CH non-thermal velocities are different compared to QS velocities. 
%Now, the non-thermal velocities cover a wide range of temperatures between 4.0 K $\leq$ Log T$_{e}$ $\geq$ 6.2 K above QS. 
We found that non-thermal velocities first rise up, attains a maximum value, and thereafter decreases further in the solar atmosphere ({\it i.e.,} quiet-Sun here). A 3$^{rd}$ order polynomial, using our QS non-thermal velocity ({\it i.e.}, Box 3) and non-thermal velocity by \cite{Teriaca1999}, has been fitted. This shows that the non-thermal velocity peaks at Log T$_{e}$$\sim$5.47 K and after that the on-average non-thermal velocity decreases in the QS.
%Using the our QS non-thermal velocity (i.e., box 3) and non-thermal velocity by \cite{Teriaca1999}, we have fitted 3$^{rd}$ order polynomial. On the basis of this 3$^{rd}$ order polynomial, we have found that $\sim$ Log T$_{e}$ 5.47 K are the non-thermal velocity peak temparature in QS. After that the on-average non-thermal velocity decrease in the QS. 
%+++++++++++++++++++++++++++++++++++++++++++++++++++++++++++++++++
\section{Discussion and Conclusions} %%%%%%%%%%%%%%
  \label{S-Discussion}
In the present work, we have studied the polar coronal hole and adjacent quiet-Sun, as observed on 10 October 2007 by {\it Hinode}/EIS.
Using {\it Hinode}/EIS spectra, a comparison between QS and CH has been performed. In the present study, we found that on average the line width of CH (39.42m \AA) is higher in comparison to the line width of QS (33.65 m\AA, {\it cf.}, Table~\ref{T-excess width}). Even, after setting a higher threshold limit of the line widths in CH ({\it i.e.}, 1.35$\times$CH average line width) in comparison to the QS threshold line width limit ({\it i.e.}, 1.20$\times$QS average width) to locate excess width locations in CH and QS, we found that CH has larger number of excess width locations in comparison to the excess width locations in QS (see Figure~\ref{fig:figure4}). We assumed that excess width locations may correspond to regions where various type of transients/explosive events may occur in solar atmosphere. Variations in the line width is directly related to the variations of the non-thermal width component (Section \ref{S-QS-CH}), therefore, the CH has higher non-thermal width as well as large number of excess non-thermal width locations in comparison to the QS. As we know that CHs are the source regions of the fast solar wind \citep{Krieger1973,Tu2005,Kohl2006, He2008,Tian2010} as well as the sites for oscillation and various type of MHD waves \citep{Banerjee2001a, Banerjee2001b,Oshea2006,Oshea2007,Banerjee2009, Gupta2010, Prasad2011}. Therefore, the presence of the larger non-thermal width and large number of excess width locations in the CH in comparison to the QS is quite obvious due to the presence of these flows/waves in the CH. 

	It is now well established that the lower temperature lines are dominated by down-flows while higher temperature lines are dominated by blue-shifts in QS and CH \citep{Chae1998,PJ1999, Teriaca1999, Tian2008a, Tian2008c, Hansteen2010, Tian2010,Neda2011}. Therefore, there is transition from red-shift to blue-shift at some temperature, which is an important observational parameter from the point of view of the TR models. Excess width regions of QS are slightly blue-shifted at Log T$_{e}$ = 5.8 K ({\it i.e.},Si~{\sc vii} 275.35~\AA), therefore, the inversion from red-shift to blue-shift has already been taken place below the Log T$_{e}$ = 5.8. On the basis of observed blue-shift (-1.9$\pm$2.0 km s$^{-1}$) of the Ne~{\sc viii} 770.428~\AA~line as seen by SUMER, which has a temperature similar to the temperature of Si~{\sc vii} line, \cite{PJ1999} have already been inferred that such transition occur at Log T$_{e}$ = 5.7 K in QS. Therefore, we can also conclude that transition ({\it i.e.}, from red-shift to blue-shift) also takes place at the same temperature in these excess width regions of the QS. Similar to excess width regions of QS, the average and excess width regions ({\it i.e.}, box, 6, 7 and 8) of CH also show marginal blue-shifts at Log T$_{e}$ = 5.8 K. Therefore, we can infer that transition temperature in CH is similar to the QS. Although, one CH box ({\it i.e.}, Box 4), which is located at the coronal hole boundary, shows strong red-shift at Log T$_{e}$ = 5.8 K. This region might have been influenced by large number of explosive events. The properties of the excess width regions of the QS are similar to CH ({\it i.e.}, excess and average width regions), but, the densities  of the excess width regions of QS are higher in comparison to the CH. In contrast, one average width box (Box 3) of QS does not show the continuous plasma up-flows while another average width box (Box 5) shows continuous plasma up-flows. Similarly, Box 3 and 5 show strong blue-shift (-3.24 km s$^{-1}$) and weak red-shift (+0.13 km s$^{-1}$) at Log T$_{e}$ = 5.8, therefore, we can not deduce a common transition temperature for these average width regions of QS. As we know that the coronal funnels as well as small and large magnetic loops are the basic building blocks of the solar atmosphere. The upper TR line blue-shifts patches are associated with coronal funnels \citep{Tu2005}, although, these blue-shift patches may be associated with the magnetic loops as well. As we have shown earlier that QS excess width boxes show continuous plasma up-flows while QS average width region does not represent the continuous plasma up flows. Similarly, the averaged densities of QS excess width regions are higher in comparison to the average width region of QS, although, the density difference is not very much. We should note that we are measuring the electron density on the basis of the intensity ratio of Fe~{\sc xii} line, which represents the electron density at coronal heights. At the coronal heights, due the large expansion of the coronal funnels, the electron density in the coronal funnels may not be different significantly from the surrounding region. %We are aiming at this small density difference between excess and average width regions provides an important signature for density concentrations. 
Therefore, slightly higher densities as well as presence of continuous plasma up-flows in the excess width regions of QS provide sufficient signature to identify that these excess width locations in QS are associated with the coronal funnels. On the basis of Doppler velocity, excess width and density, we can not distinguish the excess width boxes from the average width boxes in QSCH and CH. As we know that QS are mostly dominated by coronal loops and only a small fraction of the QS area is occupied by coronal funnels while the CH are mostly dominated by coronal funnels \citep{Peter2001}. Therefore, we have easily located the coronal funnels in QS but it is very hard to isolate  the footprints of  the coronal funnels within  coronal holes. It may be possible that all selected boxes in CH ({\it i.e.}, excess and average) are coronal funnel occupied regions because of large filling factor. %On the basis of above analysis we may assume that excess width regions of QS are CH counterpart (i.e., funnels) while the average width regions of QS do not support this fact. 
We can propose that the plasma flow inversion ({\it i.e.}, inversion from down-flows to up flows) takes place at Log T$_{e}$ = 5.7 K in CH. Similarly, the inversion of plasma flow also takes place at the same temperature (Log T$_{e}$ = 5.7 K) in excess width regions of QS. However, the average width regions of the QS do not show similar pattern of the Doppler velocity, therefore, it is hard to get conclusive plasma flow inversion temperature in these average width regions of QS. Finally, we found that excess width regions of QS show continuous plasma up-flows as well as high density concentration while the average width region of QS does not follow continuous plasma flow pattern and the density of the average width region is lower in comparison to the excess width regions of QS. Therefore, the excess width regions are the most probable locations of coronal funnels in QS, however, we could not locate the footprints of coronal funnels in the QSCH and CH using the same procedure. As we know that the coronal holes are dominated by open magnetic filed lines, therefore, it may be possible that all the selected regions of QSCH and CH (except Box 5) at these heights are within coronal funnels because in CH the parameters are similar within funnel and the adjacent background region.

 Apart from this, Doppler shift of TR and coronal lines has been also investigated while combining our results with previously reported numbers from SUMER ({\it cf.}, Figure~\ref{fig:figure6}). This new plot now covers a wide range of temperature from chromosphere up to high  corona. Doppler velocity inversion from red-shift to blue-shift takes place around Log T$_{e}$ = 5.7 K (see the peak of  4$^{th}$ order polynomial fit in Figure~\ref{fig:figure6}), which is in good agreement with  earlier results \citep{PJ1999,Teriaca1999}.  Recently Fu {\it et al}. (2014) have measured the Doppler velocity variations with temperature in the same NPCH area, however, they have focused on the on-disk plume structures. We feel that these plume structures could be also used as tracers of coronal funnels. Some of their results complement results presented here. Our focus has been on the line widths rather than the outflow velocities alone.  Earlier, 1-D models which assumes energy release in magnetic loops had successfully managed to reproduce the observed red-shift at TR temperatures and blue-shifts at coronal temperatures \citep{Teriaca1999a,Spadaro2006}. However, these models do not explicitly consider the heating mechanisms. More recently, Srivastava {\it et al}. (2014) presented  numerical model in 2-D realistic solar atmosphere, the generation of the blue-shifts (out-flows) due to heating pulses, as observed by Hinode/EIS at coronal temperature. \cite{Peter2004,Peter2006} presented, 3D models assuming that the coronal heating is caused by joule dissipation of currents produced by stressing and braiding of the magnetic fields produce red-shifts at all temperatures. Episodic injection of emerging magnetic flux, which reconnects with the existing field, produces rapid, episodic heating of the upper chromospheric plasma to coronal temperatures \citep{Hansteen2010}. % The above described models are the possible explanations of the observed red-shifts at TR temperature and blue-shifts at coronal temperatures, as we have reported here in this present case (cf.,Fig.~\ref{fig:figure6}). 
 \cite{Neda2011} measured the averaged Doppler shifts from 0.01 up to 2.1 MK and they have shown that 3D coronal models are more appropriate for the explanation of the observed Doppler velocity pattern from chromosphere up to the corona.\\   
 
The inhomogeneous solar plasma should support different MHD modes as well as acoustic wave and this has  been confirmed by observations and theory \citep{Ofman1997, Deforest1998, Banerjee2001a, Banerjee2001b, Oshea2006, Dwivedi2006, Srivastava2007, Oshea2007, Banerjee2009, Gupta2010, Prasad2011,Chem13}. It has been proposed that several mechanisms are responsible for the non-thermal broadening, {\it e.g.}, propagation of waves \citep{Mariska1978,Banerjee2009}, non-thermal motions \citep{Doschek1976, Athay1991,Chae1998}, MHD turbulence \citep{Gomez1988, Gomez1992, Hey1992} and nano-flare heating \citep{Pat2006}. In-spite of all these physical processes, the nature of the non-thermal broadening is not fully understood. We confirm here a more complete picture of the non-thermal velocity in the wide temperature range from $\sim$ 1.0$\times$10$^{4}$ K to 1.58$\times$10$^{6}$ (Figure~\ref{fig:figure8}). Initially, the non-thermal velocity increases with temperature but after certain TR temperature (Log T$_{e}$=5.8 K) non-thermal velocity decreases further upto inner corona (Log T$_{e}$=6.2 K). Undamped Alfv{\'e}n waves are responsible for non-thermal broadening \citep{Hassler1990,Banerjee1998, Wilhelm2004,Wilhelm2005,Banerjee2009}, while narrowing of spectra after a certain height in the solar atmosphere is most likely signature of the Alfv{\'e}n wave dissipation \citep{Doyle1997, Roberts2000,Pek2002,Oshea2005}. Therefore, propagation of Alfv{\'e}n waves increases the non-thermal velocity up to the inversion point in QS (Log T$_{e}$ = 5.47 K) and after the inversion point, which lies in the upper TR/lower corona, Alfv{\'e}n waves may dissipate through various mechanisms, {\it e.g.}, viscous $\&$ ohmic \citep{Roberts2000}, resonant absorption \citep{Ionson1978,Erdelyi1996,Doyle1997}, phase mixing, {\it etc}. Although, which mechanism is responsible for the Alfv{\'e}n wave dissipation can not be speculated here. This is  one possible explanation for our results related to the non-thermal velocity pattern in QS. Apart from the Alfv{\'e}n wave propagation and dissipation mechanism, another possible explanation may lie in terms of the nano-flares, which is frequently occurring at O~{\sc vi} temperature \citep{PJ1999}. \citealt{Teriaca2004} reported that non-thermal velocity peak is at the same O~{\sc vi} formation temperature ({\it i.e.}, 3$\times$10$^{5} K$) and they concluded that non-thermal velocities arise due to the prevalent occurrence of the nano-flares in this region. The presence of the lower values of non-thermal velocities above and below this region is quite obvious due to the energy loss. In the present case, the non-thermal velocity peaks at Log T$_{e}$ = 5.47 K (2.95$\times$10$^{5}$ K) in the QS. Therefore, our peak value of the non-thermal velocity is also very close to the O~{\sc vi} formation temperature and it may be probable justification for our non-thermal results. %The deposition of the energy in TR (i.e., region corresponding to O~{\sc vi} ion) by nano-flares leads to the non-thermal broadening while the dissipation of the energy above of the TR is most probable cause for the narrowing of the spectra in corona. 
%The on-average CH non-thermal velocities are higher in comparsion to the QS region (cf.,Table~\ref{table:QS_NTV} $\&$~\ref{table:CH_NTV}). As we know that CHs are the source region of the fast solar wind (e.g.,\citealt{Krieger1973,Tu2005,Kohl2006, He2008,Tian2010}) as well as the sites for oscillation and various type of MHD waves (e.g., \citealt{Banerjee2001a, Banerjee2001b,Oshea2006,Oshea2007,Banerjee2009, Gupta2010, Prasad2011}). Therefore, the presence of the larger non-thermal velocities in the CH in comparsion to the QS are quite obvious due to the presence of these flows/waves in the coronal holes. 
In conclusion, Alfv{\'e}n wave propagation and dissipation as well as prevalent occurrence of the nano-flares around at O~{\sc vi} formation temperature can explain the variation of non-thermal velocity with temperature. These two mechanisms are the viable source but that does not rule out other possibilities. Our results should help in constraining the atmospheric models. We hope that our attempt of identifying the footprints of coronal funnels based on density contrast and excess widths shed new light on the overall complexity and topology of the polar regions in general. This approach of searching for additional widths as precursors for transients can be further improved with future better resolution spectrographs.     
%\section{Conclusions} %%%%%%%%%%%%%%%%%%%%%%%%%%%%%%%%%%%%%%%%
      %\label{S-Conclusion} 
%On the basis of Hinode/EIS spectroscopic observation in a North Polar Coronal Hole, we have made a comparison between QS and CH.
  %We have found that CH has higher line width as well as larger number of excess width locations in comparison to the QS. 
%The presence of the larger line width as well as larger number of excess width pixels location in CH  confirms that CH is the site of various type of waves and oscillations. 

%%Apart from this, we have measured the non-thermal velocities from upper TR (Log T$_{e}$ = 5.8 K) up to the corona (Log T$_{e}$ = 6.2 K) in QS and CH. After including previous relevant results  with our results, we  found that for QS, non-thermal velocity increases with temperature but after a certain temperature (Log T$_{e}$ = 5.7) the non-thermal velocities decrease in QS. 
%%%%%%%%%%%%%%%%%%%%%%%%%%%%%%%%%%%%%%%%%%%%%%%%%%%%%%%%%%%%%%%%%%%%%%%%%%%
\begin{acks}
 We acknowledge the Hinode/EIS observation for this study. Hinode is Japanese mission developed and launched by ISAS/JAXA, with NAOJ as domestic partner and NASA and UKSA as international partners. P. Kayshap acknowledges the support from Indian Institute of Astrophysics, Bangalore for his visit.  
\end{acks}
%%%%%%%%%%%%%%%%%%%%%%%%%%%%%%%%%%%%%%%%%%%%%%%%%%%%%%%%%%%%%%%%%%%%%%%%%%%

\end{article} 
\end{document}